\documentclass[journal,10pt,twocolumn]{IEEEtran}
\usepackage{graphicx, times, amsmath, amsfonts, comment,amsthm}
\usepackage[noend]{algorithmic}
\usepackage{algorithm, array}
\usepackage{epstopdf}
\usepackage{verbatim}
\usepackage{tikz}
\usetikzlibrary{trees}

\newcommand{\beq}{\begin{equation}}
\newcommand{\eeq}{\end{equation}}

\theoremstyle{plain}

\theoremstyle{plain}

\theoremstyle{plain}

\theoremstyle{plain}

\theoremstyle{plain}
\newtheorem{defcounter}{Definition}
\newtheorem{definition}[defcounter]{Definition}

\newcommand{\figref}[1]{Fig.~\ref{#1}}

\begin{document}
\title{Mobility-Aware  Analysis of 5G and B5G Cellular Networks: A Tutorial}
\author{ Hina Tabassum, Mohammad Salehi, and Ekram Hossain\thanks{The authors are with the Department of Electrical and Computer Engineering at the University of Manitoba, Canada
 (emails:  \{Hina.Tabassum, Ekram.Hossain\}@umanitoba.ca, salehim@myumanitoba.ca). 
 }}


\maketitle

\begin{abstract}


Providing  network connectivity to mobile users is a key requirement for cellular wireless networks. User mobility impacts network performance as well as user  perceived  service  quality. For efficient network dimensioning and optimization, it is therefore required to characterize the mobility-aware network performance metrics such as the handoff rate, handoff probability, sojourn time, direction switch rate, and users' throughput or coverage. This characterization is particularly challenging for heterogeneous, dense/ultra-dense, and random cellular networks such as the emerging 5G and beyond 5G (B5G) networks. In this article, we provide a tutorial on mobility-aware performance analysis of both the spatially random and non-random, single-tier and multi-tier cellular networks. We first provide a summary of the different mobility models which include purely random models (e.g., random walk, random way point, random direction), spatially correlated (e.g., pursue mobility, column mobility), and temporally correlated models  (e.g., Gauss-Markov, Levy flight). The differences among various  mobility models, their statistical properties, and their pros and cons are presented. We then describe two main analytical approaches (referred to as {\em trajectory-based} and {\em association/handoff based} approaches) for mobility-aware performance analysis of both random and non-random cellular networks. For the first approach (which is more general but less tractable than the other approach),  we describe a general methodology and present several case studies for different cellular network tessellations  such as square lattice, hexagon lattice, single-tier and multi-tier models in which base-stations (BSs) follow a homogeneous Poisson Point Process (PPP). For the second approach, we also outline the general methodology. In addition, we discuss some limitations/imperfections of the existing techniques and provide  corrections to these imperfections. For both the approaches, we  present selected numerical and simulation results to calibrate the achievable handoff rate and coverage probability in various network settings. Finally, we point out specific 5G application scenarios where the impact of mobility would be significant and outline the challenges associated with mobility-aware analysis of those scenarios.  

\end{abstract}


\section{Introduction}

\subsection{5G/B5G Networks and User/Device Mobility}

The fifth-generation (5G) mobile communication networks are envisioned to support massive connectivity (millions of devices per sq. km), higher data rates, lower transmission delays  (around 5 ms) in user plane  and  (around 10 ms) for control plane, and devices with very high mobility speeds ($\sim$ 500 kmph)~\cite{5g,5g2016}. 5G networks will support ultra-reliable low latency communication  (URLLC), enhanced Mobile BroadBand (eMBB) communication and massive machine type communications (mMTC) for a wide  variety of applications  such as augmented/virtual reality, ultra-high-definition video, cloud storage, Internet of Things (IoT), Internet of Vehicles (IoV), smart home, and smart cities. In the sequel, 5G networks
will  utilize  ultra-dense deployment of access points, higher  frequency  bands (e.g., mm-wave, free-space optics [FSO], visible light, and Tera Hertz)  via carrier aggregation or dual connectivity, and massive antennas to overcome higher path loss and blocking associated with such high frequencies.
Further, technologies enabling
device-to-device communications (D2D), cognitive radios, inter-vehicular (V2V), vehicle-to-pedestrian (V2X), vehicle-to-infrastructure (V2I), drone-to-infrastructure (D2I), and drone-to-user (D2X) communications are expected to be integral parts of  future 5G/B5G wireless networks.  The key features of 5G/B5G cellular networks include spatial  randomness of network tessellation, heterogeneity of base-stations (BSs), dense/ultra-dense nature of network deployment, and diversified mobility patterns of users/devices and network nodes.

Modeling and analysis of user mobility plays a vital role in optimizing the design and performance of cellular wireless networks.  User mobility directly impacts the following:
\begin{itemize}
\item {\em Resource management aspects} such as  channel  allocation  schemes,   multiple   access   mechanisms, estimate of network  capacity,  call blocking rate, traffic  volume  per  cell, users' quality  of  service  (QoS),  signaling  and  traffic  load  estimation, etc.  
\item {\em Radio propagation aspects} such as  signal  strength  variation,  interference level,  call dropping rate, handoff\footnote{An event where a mobile user connected to one BS switches to another BS in order to maintain its ongoing transmission and/or network connectivity.}  algorithms (typically based  on signal  strengths).  
\item {\em Location management aspects} that include  location   area   planning,   multiple-step   paging strategies,  data  location  strategies,  database  query  load.  
\end{itemize}
The significance of mobility-aware network performance analysis is thus evident in optimizing the interference and network-wide resource management  and accordingly user QoS provisioning through network dimensioning, optimizing handoff thresholds, and handoff methods. Since mobility directly affects the availability of network resources, 
mobility-aware performance modeling and optimization of systems such as device-to-device (D2D) communications, cognitive radio networks, unmanned aerial vehicles (UAV) or drone networks, and vehicle communications  is of immediate relevance.

Due to the irregular deployment of BSs in emerging 5G/B5G cellular networks, mobility-aware performance analysis must capture the location uncertainty of the BSs~\cite{adve19, adve20}. One of the common and analytically tractable approaches to incorporate spatial randomness in cellular networks is to model BSs with the homogeneous Poisson point process (PPP) where the number of nodes in area $|A|$ is a Poisson random variable and the  nodes are uniformly distributed. The accuracy of PPP-based model for a two-tier cellular network was examined in \cite{dhillon2012modeling} and it was shown that the signal-to-interference-plus-noise ratio (SINR) coverage probability (i.e., the probability that the instantaneous {\rm SINR} is above a given target threshold) of a user in a  4G network lies between the PPP model (pessimistic lower bound) and the  hexagonal
grid model (optimistic upper bound).  Since then there have been various follow-up research studies where PPP-based modeling has been used to assess the rate and coverage probability of  users~\cite{mimo,zhang2014stochastic,vlc}.
Taking into account the spatial randomness of cellular networks using stochastic geometry tools, some of the pioneering research studies in which  mobility-aware performance analysis has been carried out are \cite{lin2013towards,bao2015stochastic,adve2015handoff,shin2017equivalent}. 

Due to inherent heterogeneity of 5G/B5G cellular networks, a given   user may traverse   among  extremely diverse networks and cells with diverse transmission frequencies, e.g., radio frequency cells and mm-wave cells. This will result in frequent handoffs that create heavy signaling overhead and deteriorate user's experience even at low to moderate velocities. Mobility modeling will therefore be more important  compared to homogeneous scenarios where users move among the same type of cells. Note that compared to horizontal handoffs (where a user switches to another BS with similar transmission features), vertical handoffs (where a user switches to another BS having diverse transmission characteristics such as transmit power and frequency) may impact the overall system more adversely. The reasons include extra control signaling, possible call drops due to channel unavailability, and varying channel attenuation factors due to different transmission frequencies.  Note that a mobile user may improve data rate by performing a vertical handoff to a more dense tier, but this may also result in frequent handoffs/delays, which potentially deteriorates the service quality. Balancing the trade-offs  among handoff rate, service delay, and achievable coverage/data rate in heterogeneous, dense, and random 5G/B5G cellular networks is therefore an open challenge.


\subsection{Mobility and Handoff Management}

Mobility of users/devices results in handoff.  The number of handoffs   are proportional to the intensity of BSs and velocity of users. The handoff process requires a smooth transfer of a connected user when moving from one cell to another with the guaranteed QoS. The objective of efficient handoff/mobility management is to reduce radio link failures during handoff, handoff failures, and ping pong events.   Mobility management was included in the first release of  Long-Term Evolution (LTE) standard (Rel-8) for homogeneous networks \cite{lterel8} where the handoffs are generally based on the measurement of signal strengths from the neighboring BSs and are affected by the time/frequency selectivity of the  propagation channel. The condition for handoff can be  written as:
\begin{equation}\label{1}
S_{\mathrm{neighboring}}-S_{\mathrm{serving}} \geq O_s -O_n + H_s +H_{\mathrm{off}}
\end{equation}
where $O_s$ and $O_n$ represent offsets   of   serving   and   neighboring   BSs  configured  by  the  network operator, respectively. $H_s$ is  the  handoff hysteresis parameter which corresponds to the traffic load of different BSs and $H_{\mathrm{off}}$ is a parameter specific for each pair of network BSs. handoff occurs when the condition in \eqref{1} has  been  met  for a specific  duration, i.e., time  equal  or  longer  than  value  of  the  Time-To-Trigger (TTT)\footnote{This is the minimum time for which the handoff criterion needs to hold.}.  Later, mobility enhancements for co-channel heterogeneous networks are considered in LTE Rel-11 \cite{lterel11}.  Specifically, handoff procedures are optimized by dynamically adapting handoff parameters for different cell sizes and user velocities.

\begin{figure*}[!ht]
\begin{center}
\includegraphics[width=5in, height=4.75in]{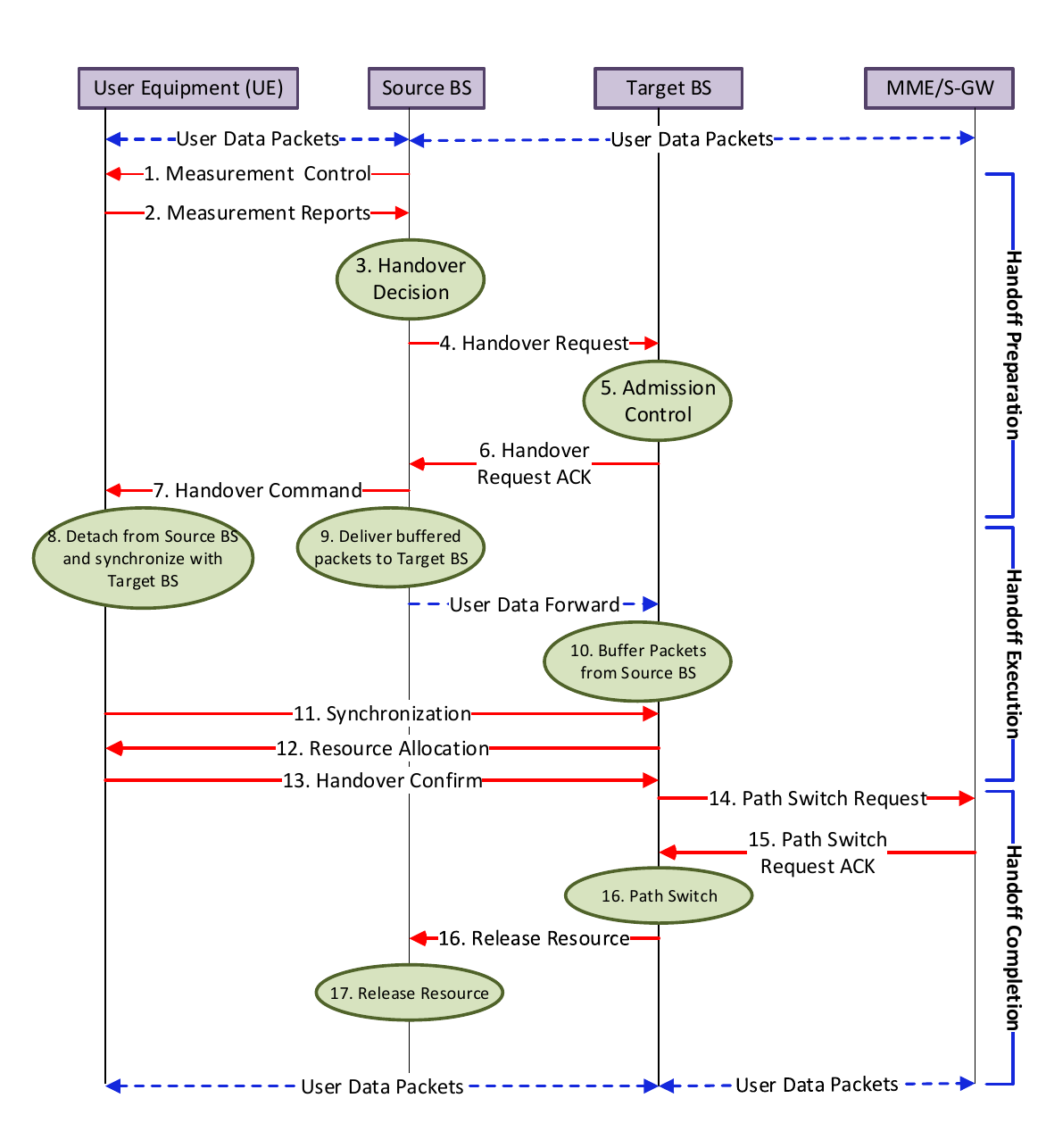}
\end{center}
\caption{Procedure for handoff in 3GPP LTE \cite{lte}. }
\label{lte}
\end{figure*}

The  LTE network  architecture  is composed of BSs  (providing  both  user  plane  and  control  plane  to  users),   mobility management entity (MME) and system architecture evolution gateway (S-GW)~\cite{lte6}. 
BSs are connected to MME/S-GW by the S1 interface and connects to each other through X2 interface. The S-GW controls inter-3GPP  mobility  while  routing  and  forwarding  the  user data packets.  Note that the user mobility support is needed whether the user is in idle mode or in connected mode \cite{lte7}. When a user turns on Public Land Mobile Network (PLMN) is selected and the user searches for a suitable cell of selected PLMN and  tunes to 
its control channel. This procedure is referred as ``camping 
on the cell". 
In connected mode, LTE utilizes a network controlled and user-assisted handoff procedure.  The  LTE handoff procedure is shown in \figref{lte} and the steps  are summarized as follows~\cite{lte}:
\begin{itemize}
\item Each user continues to measure the received signal strength $S$ from the serving and neighboring BSs. 
\item To initiate handoff, the user reports the measurements (e.g., reference   signal   received   power   (RSRP)   and reference  signal  received  quality  (RSRQ)) taken from neighboring
BSs to their respective serving BS. 
\item  {\em Handoff preparation}: The serving BS makes the handoff decision based on the measurement reports and radio resource management information of the target BS. 
\item The serving BS then sends the handoff request to the target BS. Based on the admission control of the target BS, the serving BS gets the acknowledgment from the target BS. Once the serving BS receives the acknowledgment, it transfers all information to the user.
\item {\em Handoff execution}: The user then sends a confirmation signal to the target BS. After that the target BS
sends the path switch command to MME/S-GW.
\item {\em Handoff completion}: After path switch completion, user  releases  the  serving  BS resources. 
and  access  the  target  BS  using  the  random  access  channel (RACH). Upon  synchronization  with  the  target  BS,  the  user
sends the confirmation message to notify the network that the
handoff has been completed.
\end{itemize}
It can be observed that the signaling overhead (due to the searching  process and  handoff related signaling between   the   user,   serving   BS,   target BS,  and  core  network) can increase the {\em handoff delay}\footnote{The handoff delay is measured  from the  beginning  of  initiation  phase  to  the  end  of  execution phase.} significantly.  

\subsection{Mobility-Aware Performance Measures}


Network performance metrics such as coverage or throughput need to incorporate the impact of user/device/node mobility and network performance analysis methodologies need to be tractable. Some relevant mobility-aware  network performance metrics include~\cite{lin2013towards,adve2015handoff,bao2015stochastic,hesham2017velocity}:
\begin{itemize}
\item {\em Handoff rate}~\cite{lin2013towards, bao2015stochastic}: This is given by the expected number of handoffs divided by the average transition time taken by a user to move from one way point to another.
\item {\em Sojourn time (or dwell time)}~\cite{lin2013towards}:  This refers to the time a mobile node resides in a typical cell. In other words, this is the time for which a BS provides service to a node.
\item {\em Direction switch rate}~\cite{lin2013towards}: This is the reciprocal of the sum of transition time and pause time.
\item {\em Handoff probability}~\cite{adve2015handoff}: 
{This is the probability that the user crosses over to the neighboring cell in one movement period} (i.e., the probability that the serving BS does not
remain the best candidate in one movement period). 
\end{itemize}
{By definition, the handoff rate is the average number of handoffs per unit time, i.e., 
\begin{IEEEeqnarray}{rCl}
	H &=& \mathbb{E} \left[ \text{Number of handoffs per unit time} \right] \nonumber \\ 
	  &=& \sum_{k=1}^{\infty} k \mathbb{P} \left( \text{Number of handoffs per unit time} = k \right). \nonumber
\end{IEEEeqnarray}}
{For low velocities, the handoff rate is equal to the probability of the handoff since the number of handoffs in a unit time is one with probability $\mathbb{P}(H)$ and zero with probability $\mathbb{P}(\bar{H})=1-\mathbb{P}(H)$, i.e.,
$H \sim \mathbb{P}(H),
$
where $\mathbb{P}(H)$ denotes the probability of handoff. Also, when the BS density is low, the handoff rate can be approximated by the probability of handoff for larger range of velocities. }
Note that handoff rate is inversely proportional to the expected sojourn time; however, their distributions can be quite different~\cite{lin2013towards}.  

Generally,  it is difficult to incorporate the impact of handoff rate and sojourn time  on the rate or coverage probability of a typical mobile user. This is why the coverage or rate expressions  are typically derived for  stationary (but randomly located) users and spatial averaging is then performed.  
Nevertheless, due to heterogeneous and ultra-dense nature of 5G/B5G networks, it is not sufficient to compute the coverage and rate  metrics only for stationary users.  The reason is that the optimal association of a  user  (from the perspective of data rate maximization) may not remain optimal due to higher handoff rates. As such, the trade-off between handoff rate and data rate needs to be captured and the performance measures should be designed accordingly.
In this regard, the relevant mobility-aware network performance metrics include:
\begin{itemize}
\item{\em Mobility-aware coverage probability}~\cite{adve2015handoff}: can be defined as a sum of (i) probability of the joint event that the user is in coverage and no handoff occurs and (ii) probability of the joint
event that the user is in coverage and handoff occurs penalized by the cost of handoff.
\item{\em Mobility-aware throughput}~\cite{heshamletter}: is defined as   the  traditional spatially averaged throughput of a user multiplied by a factor $(1- H d)$  where $H$ is the handoff rate and $d$ is the  delay per handoff. This allows to incorporate the impact of handoff on the users' achievable throughput.
\end{itemize}

\subsection{Scope of the Tutorial}


This tutorial provides a comprehensive review and comparative analysis of tractable analytical methodologies presented in~\cite{lin2013towards,bao2015stochastic,adve2015handoff,shin2017equivalent} for mobility-aware performance analysis  (in {\em layer 2}) of emerging 5G/B5G cellular networks. {\em The issues related to handoff management (e.g., optimization of the handoff parameters, resource allocation for handoff) and the analysis of aspects related to radio propagation (e.g., signal  strength  variation,  time  dispersion of signals) are not within the scope of this tutorial}. 

We will first review different mobility models that can potentially mimic the movement patterns of users/devices and wireless nodes such as air crafts, high speed trains, vehicles, wearables, drones, unmanned air vehicles (UAVs) etc. These models include purely random models (e.g., random walk, random way point, random direction), spatially correlated (e.g., pursue mobility, column mobility), and temporally correlated models  (e.g., Gauss-Markov, Levy flight). The distinctive features of the aforementioned  mobility models, their statistical properties, and their benefits and drawbacks will be presented. Then we will provide a summary of the existing state-of-the-art of mobility and handoff analysis (sheer majority of which are for {\em spatially non-random} cellular networks) based on simulation and theoretical approaches. 

\begin{figure}[t]
\begin{center}
\includegraphics[scale=.65]{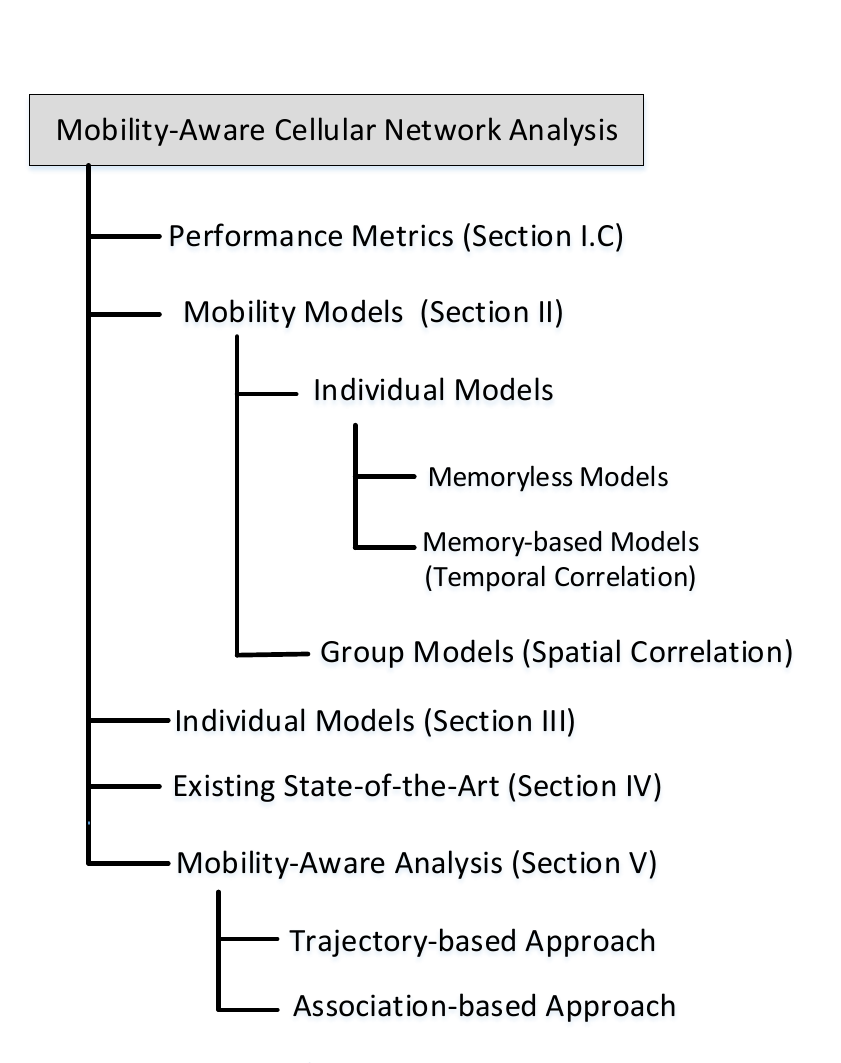}
\end{center}
\caption{Organization of the tutorial.}
\label{intro}
\end{figure}

We will then provide a systematic introduction to the existing analytical methodologies for mobility-aware performance analysis in  {\em spatially random} cellular networks. These methodologies are general to accommodate  a variety of mobility models to conduct mobility-aware performance analysis. 
In this tutorial, two major approaches, namely, the {\em trajectory-based} and the {\em association-based} approaches, are described. For the trajectory-based approach, which is more general (however less tractable), we highlight a general methodology to perform mobility-aware performance analysis for both random and non-random cellular networks.
Case studies are presented for different cellular network topologies  such as square lattice, hexagon lattice, single-tier and multi-tier models in which BSs follow a homogeneous PPP.  For the association-based approach, we will also outline the general methodology to calculate the handoff probability and the mobility-aware coverage probability. In addition, some limitations/imperfections of the existing techniques in this approach will be pointed out and corrections to these will be also provided. Also, for both the approaches, we will present selected numerical and simulation results to calibrate the achievable handoff rate and coverage probability  by a user in various network settings. Finally, we will outline some research directions and potential approaches for mobility-aware analysis of 5G/B5G networks. The organization of the article is shown in \figref{intro}.

\begin{figure*}[t]
\begin{center}
\includegraphics[scale=.8]{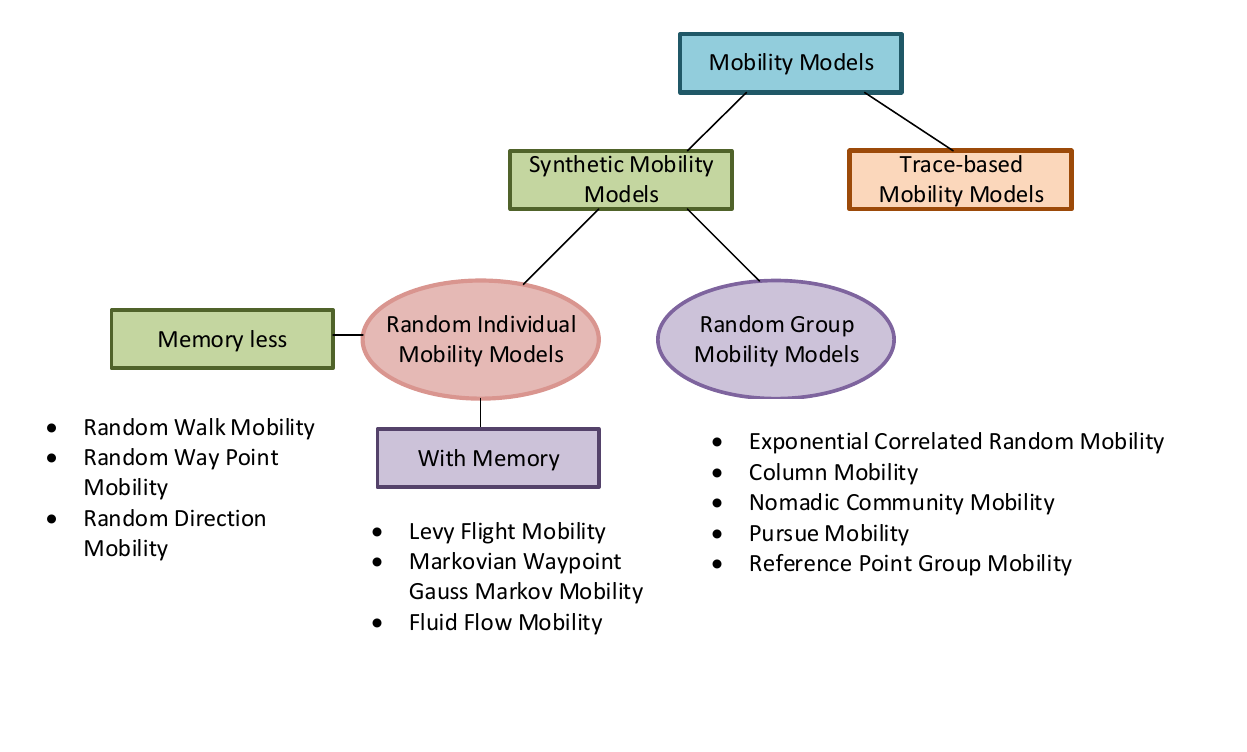}
\end{center}
\caption{Classification of various mobility models potentially applicable to cellular network modeling, analysis, and optimization. }
\label{fig1}
\end{figure*}

\section{Classification of Mobility  Models}

The handoff delay and signaling overhead may become significant  in ultra-dense networks given that each small cell may receive a large number of handoff requests, followed by the execution of  the admission control algorithm for each accepted request. Some of these handoff requests may even be unnecessary, especially for truly mobile users who are expected to enter and leave the cells swiftly. Therefore, sophisticated  mobility-aware handoff procedures (based on the information of users such as their location, speed, and direction) will be required. Subsequently, characterizing the statistics of  the location, speed, and direction of users is of prime relevance since it will enable us understanding the mobility pattern\footnote{Often, the mobility models and mobility patterns are used synonymously.
However, one must carefully distinguish between them since the mobility patterns can be obtained by tracking real moving objects (e.g., pedestrians, vehicles, aerial, robot, and outer
space motion~\cite{schindelhauer2006mobility}), while mobility models offer mathematical formulations for different mobility patterns.} of users, deriving sophisticated handoff criteria, and mobility-aware performance analysis of  cellular networks.  We  will review various mobility models and their potential applications in emerging 5G/B5G cellular networks. 

The precision of mobility models can be measured in terms of how close they can model the real mobility patterns of users and/or different kind of wireless nodes.
However, attaining this precision may result in  huge computational cost or reduced mathematical tractability. The mobility models  to characterize the movement patterns of mobile nodes in wireless networks are typically classified as follows~\cite{survey2006mobility}:
\begin{itemize}
\item {\em Trace-Based Mobility Models}: 
are obtained by measurements of deployed systems (e.g., from logs of connectivity or location information of mobile users)~\cite{trace1, trace2,trace3}. These models are realistic in terms of the movement and topology of the area, e.g., CRAWDAD project~\cite{crawdad}.  These traces are valuable  for the performance assessment and optimization 
of handoff protocols but may not serve as benchmarks for the scientific
community. The reason is that the available real traces may not be applicable and generalized for a variety of scenarios.
\item {\em Random Synthetic Models}:  are mathematical models to characterize the movement of the devices. The models are generally simple and analytically tractable, but may not reflect realistic mobility patterns~\cite{survey1}. Examples include Bayesian models that are capable of mimicking random behavior of a node (or group of nodes), Brownian motion \cite{brownian} that characterizes the diffusion of tiny particles with a mean
flight time and a mean pause time between flights and
the Levy pattern \cite{levy} which is more diffusive than Brownian
motion  and is a good approximation
of human walk in outdoor environments.
\end{itemize}
Due to the differences in the trace acquiring methods, sizes of trace
data, and data filtration techniques, a trace-based mobility model for
one network data set may not be applicable to other network
scenarios. 
The traces may not be publicly available.
The available traces may not be sufficient to analyze the network performance as the parameters  such as speed or the density of the nodes cannot be varied.
Nevertheless, such models are precise and realistic for a specific scenario.  On the other hand, random mobility models are generic and mathematically tractable. 
Subsequently, random mobility models can be used for rapid  assessment, mathematical analysis, and optimization of a variety of network scenarios~\cite{synthetic}.

%
%
%
%
%

The random synthetic models can be further classified as:

\begin{itemize}
\item {\em Individual Mobility - Memoryless}: In the individual memoryless mobility models, a mobile user moves independent of other nodes. The location, speed, and movement  direction   of  a given mobile   node   are   neither   affected   by other   nodes   in   its vicinity nor a function of its  previous velocities and locations. These models are mathematically tractable but may not be close to reality. For example,  to avoid collision on a road, the  speed  of  a  vehicle  cannot  exceed  the  speed  of  the  vehicle  ahead  of  it.   Therefore,  it is evident that the mobility of users could   be   influenced   by other  neighboring  nodes which results in spatial correlation or spatial dependency among mobile nodes~\cite{bai2004mathsurvey}.  Furthermore, these models are vulnerable to sudden  stops,  sudden  acceleration,  and  sharp  turns. Typically, the   velocity   of   vehicles   and   pedestrians     accelerate   incrementally rather than randomly and the direction changes are also smooth leading to temporal correlation or temporal dependency among the mobility parameters.

\item {\em Individual Mobility - With Memory}: In the individual mobility models with memory, a mobile user moves independent of other nodes. Different from the memoryless models, a node's next location is a function  of its past locations and velocities. These models are also referred to as mobility models with temporal dependency.

\item {\em Group Mobility:} The group-mobility models are generally an extension of the individual mobility models. These models either utilize a mathematical function to describe the mobility behavior of a group (e.g., exponential correlated random mobility model~\cite{survey1}, community model, and column mobility model~\cite{sanchez2001anejos}). These models are also referred to as mobility models with spatial dependency.
The set of mobile nodes in column mobility model form a line and move forward in a particular direction. A community mobility model is the one where a set
of mobile nodes move together from one location to another. 
Another type of group mobility models tends to mimic the behavior of mobile nodes that associate with a group leader (e.g., pursue mobility model~\cite{sanchez2001anejos}, reference point group mobility model~\cite{RPGM}). The pursue model allows the users in a group to follow a target node moving over the simulation area. The reference point group mobility model considers the group movement  based upon
the path traveled by a logical center according
to an individual mobility model described before~\cite{RPGM}. 

\end{itemize}

A classification of the useful mobility models is provided in \figref{fig1} and a more extensive description of some of these models can be found in  \cite{survey1} and \cite{survey2}. The mobility models with spatial and temporal dependencies have not been exploited fully in the context of cellular networks. {\em To date, most of the mobility-aware performance analysis is based on memoryless models such as random walk or RWP models}.  Note that group-based mobility models can also be very relevant for vehicular applications since high-speed trains, aircrafts, or vehicles will require group handoffs since a group of people will be transitioning from one BS to another BS.  Moreover, incorporating the impact of temporal correlations due to human walking tendency and human clustering behaviors is another important direction to be considered.

\section{Random Synthetic Mobility Models}
In this section, we provide an overview and taxonomy of  random synthetic individual mobility models as they are relatively tractable, and hence, beneficial for rapid performance modeling and assessment of mobile users in various 5G cellular network scenarios.   The goal is to provide readers with a fundamental background to easily understand and compare  different statistical models and eventually identify the one according to their requirements.

Consider mobile users move randomly in a two-dimensional spatial region. The random movement conducted by each mobile user can be expressed by  stochastic sequence $S$ as:
\begin{equation}
S = (L,\Theta , T , T_p)
\end{equation}
where a mobile user covers a distance of length $L$ (the distance is generally referred to as flight length in the literature)  and flight duration $T$ followed by a pause of duration $T_p$. $\Theta$ represents the direction
change from the previous flight. 
Before taking another flight, a  mobile user chooses the variables $(L,\Theta, T , T_p)$ according to their respective distributions.
The mobile velocity can then be expressed
as $V = L/T$. The beginning of each transition is
referred to as a way point, whereas the trajectory of a mobile node  can be expressed as a set of way points and lines generated
by $S$. 

\begin{table*}[!ht]
\centering
\caption{Features of Individual Memoryless Random Mobility Models}
\label{my-label}
\begin{tabular}{|p{1.3cm}|p{2.7cm}|p{2.3cm}|p{1.75cm}|p{2.7cm}|p{2.2cm}|p{2.2cm}|}
\hline
 {\bf Mobility Model} & { \bf Flight Length ($L$)} & {\bf  Direction ($\Theta$) PDF}  & {\bf Flight Time ($ T$)}  &  {\bf  Pause Time ($T_p$)} & {\bf Limitations}  & {\bf Benefits}
\vspace{3mm}
\\ 
\hline
Random Walk~\cite{pearson1905problem}   & Eq.~\eqref{brownian1} & $\frac{1}{2\pi}, \quad 0 \leq \Theta \leq 2\pi$ & $T = $ constant  & $T_p$ = 0 
&Border effect, Far from human mobility& Tractable\vspace{5mm}
\\
\hline
Random Way point~\cite{johnson1996dynamic}  & PDF: Eq.~\eqref{rwp} - rectangular, Eq.~\eqref{circ} - circular & \cite[Eq. 13]{bai2004mathsurvey},  Eq.~\eqref{theta} & PDF: Eq.~\eqref{tf}  & $T_p = $ constant 
&Non-uniform $\Theta$ and node distribution & Tractable, Available in network simulators  NS-2 and NS-3
\vspace{3mm}
\\
\hline
Modified Random Direction~\cite{lin2013towards} & PDF: $2 \pi \lambda l e^{\lambda \pi l^2}$ &  $\frac{1}{2\pi}, \quad 0 \leq \Theta \leq 2\pi$ & PDF: Eq.~\eqref{tf1} and Eq.~\eqref{tf2}   & $T_p =$ constant 
& Way points are not i.i.d and form a Markov process & Tractable, Uniform node distribution, Close to Levy Flight
\vspace{3mm}
\\
\hline
Truncated Levy~\cite{levy}   & $\frac{1}{2\pi}\int_{-\infty}^\infty e^{-i x l-|c l|^{\alpha}}  dx$ &$\frac{1}{2\pi}, \quad 0 \leq \Theta \leq 2\pi$  & $k L^{1-\rho}$&$\frac{1}{2\pi}\int_{-\infty}^\infty e^{-i x t-|c t|^{\beta}}  dx$ & Accuracy (within 10 km)& Close to human mobility   
\\
\hline
\end{tabular}
\end{table*}

\subsection{Memoryless Individual Mobility Models}
The memoryless mobility models retain no knowledge about the previous locations and velocities of the mobile nodes. This characteristic limits the practicality of the memoryless models to some extent because mobile nodes typically have a predefined destination and speed which may affect future destinations and speeds. Some of the popular  memoryless individual mobility models  are reviewed  in the following~\cite{survey2006mobility}.

\vspace{0.2cm}
\subsubsection{ Random Walk (Brownian) Model~\cite{pearson1905problem,brownian}} 
The concept of random walk was first invented by Pearson in 1905~\cite{pearson1905problem}. In this model, a user randomly chooses a direction $\Theta$ and speed $V$ from a given range and then travels for a fixed duration $T$ or, equivalently, a distance $L=V T$. 
The pause time of the mobile node $T_p=0$.
For every new interval, mobile node randomly and uniformly chooses  direction in the range  $\Theta \in [0 \quad 2\pi]$. The  speed follows a uniform
distribution or a Gaussian distribution and is in the range 
 $V \in [v_{\mathrm{min}}  \quad v_{\mathrm{max}}]$. The PDF of $L$ can be given as:
\begin{equation}\label{brownian1}
f_L(l)=\frac{\frac{1}{T \sigma_v} \phi\left(\frac{l-T \mathbb{E}[v]}{T \sigma_v}\right)}{\Phi(\frac{v_{\mathrm{max}}-\mathbb{E}(v)}{\sigma_v})-\Phi(\frac{v_{\mathrm{min}}-\mathbb{E}(v)}{\sigma_v})}
\end{equation}
where $\phi$ and $\Phi$ denote, respectively, the probability density function (PDF) and the cumulative distribution function (CDF) of standard Gaussian random variables.
During the flight, the node moves with velocity vector $[V \mathrm{cos}\Theta \quad V \mathrm{sin}\Theta]$.
At boundaries, the user can either
reflect back from the boundary, or wrap around   resulting in sharp directional changes termed as {\em Border Effect}~\cite{bai2004mathsurvey}.

Random walk   was  used  in  various  research studies  \cite{rwp4,rwp5}. For example, \cite{rwp4} used the random movement  to derive the mean cell sojourn time and  \cite{rwp5} conducted a systematic tracking of
the random movement of a mobile node. At each instant, the  area is partitioned into several regions according to previous, current, and next direction of the mobile node~\cite{rwp5}. Explicit mathematical  conditions for  movements from the current region into the next region were derived. Based on these conditions, channel holding time, cell residence time, and the number of handoffs were derived.  Although this is  one of the most popular mobility models,  it fails to capture realistic human/vehicle movement patterns. 

\vspace{0.2cm}
\subsubsection{Random Way point (RWP) Model~\cite{johnson1996dynamic,hyytia2006spatial}} 
The RWP model was first introduced in 1996~\cite {johnson1996dynamic} to model the movement of mobile users. RWP is an extension of  random walk mobility model with pause  time  between  changes  in  direction  or  speed. RWP  has  a  simple  implementation  and   is  available  in
typical network  simulators such as in  NS-2 and NS-3.  
In this model, each node picks a random destination (referred to as way point) uniformly distributed within a finite space and travels with a speed uniformly chosen from an interval $V \in [0 \quad V_{\mathrm{max}}]$. 
Then the user moves along the line (whose length is called transition
length $L$) connecting its current way point to the newly selected
way point at a chosen velocity $V$. 
Upon reaching the destination, the process repeats itself (possibly after a random pause time $T_p$). 

\cite{bettstetter2004stochastic}  described  RWP as a discrete-time stochastic process. It was shown that the average flight length in a 
single  epoch  over  all  mobile nodes (i.e., spatial average) is equal to the average flight length of a specific node  over time (i.e., time average). As such, RWP  has mean-ergodic property according to the  theory of  random processes.  Once RWP is  ergodic,  the PDF of flight length $L$ can be described by considering the Euclidean distance between two independent random
points in the simulation space. Consequently, by applying the standard geometrical probability theory, the PDF  of   transition (or flight)  length  $f_{L}(l)$   in a rectangular region (of length  $a$ and  width  $b$) can be given  as follows~\cite{bettstetter2004stochastic}:
\begin{equation}
f_{L}(l)= \frac{4 l}{a^2 b^2} f_0(l)
\label{rwp}
\end{equation}
where
\begin{equation}
f_0(l)=
\begin{cases}
\frac{\pi}{2} a b - a l -b l+\frac{l^2}{2}  & 0 \leq l \leq b\\
G(l) &b < l<a\\
G(l) - H(l) & a < l<\sqrt{a^2+b^2}
\end{cases}
\end{equation}
in which $G(l)=ab \; \mathrm{sin}^{-1}\frac{b}{l} + a \sqrt{l^2-b^2} -\frac{b^2}{2} -a l$ and $H(l)=ab \; \mathrm{cos}^{-1}\frac{a}{l} + b \sqrt{l^2-a^2} -\frac{l^2+a^2}{2}$.
The PDF of $L$ for a circular region of radius $a$ can be derived as follows~\cite{bettstetter2004stochastic}:
\begin{equation}\label{circ}
f_{L}(l)= \frac{8 l}{2 \pi a^2} \mathrm{cos}^{-1}\frac{l}{2 a} -\frac{l}{2 a}\sqrt{1-\left(\frac{l}{2 a}\right)^2}.
\end{equation}
The PDF of flight time can then be derived as follows:
\begin{equation}\label{tf}
f_{T}(T)=\int_{v_{\mathrm{min}}}^{v_{\mathrm{max}}} v f_L(v t) f_V(v) dv.
\end{equation}
It was observed in \cite{bettstetter2001mobility,blough2004statistical,bettstetter2004stochastic}
that the spatial  node   distribution  becomes non-uniform   with the increasing simulation runs.  As such, the node density becomes maximum  in  the  center  and nearly zero at the boundary of the simulation space~\cite{navidi2004stationary}. This phenomenon happens since the nodes are likely to either  move  towards  the  center  of  simulation  field  or  choose  a  destination  that  requires  movement  through  the  middle~\cite{bai2004mathsurvey}. 
\cite{bettstetter2001mobility} showed  that  the  underlying reason  
is  the  non-uniform distribution  of  the  direction  angle  at  the  beginning  
of  each  flight.  Based on the PDF of direction angle $\Theta$,
\begin{equation}\label{theta}
f_\Theta (\Theta)=\int_0^{2\pi} \int_0^a f_\Theta (\Theta | r) \frac{r}{\pi a^2} dr d\phi,
\end{equation}
it was shown that the  probability of moving in a direction towards the boundary within interval $[\frac{\pi}{2} \quad  \frac{3 \pi }{2}]$ is only 12.5\%. However, the node moves toward the center within interval $[-\frac{\pi}{4}  \quad\frac{ \pi }{4}]$ with a probability of 61.4\%~\cite{bettstetter2001mobility}.

\vspace{0.2cm}
\subsubsection {Random Direction Model~\cite{royer2001analysis}} 
This model was constructed mainly to address the tractability issues arising due to  non-uniform node distribution of RWP \cite{royer2001analysis}. In this
model, at each way point the mobile node chooses a random
direction $\Theta$ uniformly distributed on $[0, 2\pi]$ and  a velocity $V$ from some distribution. 
Then the node moves to the next way point (determined
by transition length and direction) at the chosen velocity. 
After  the  node  reaches  the  boundary of  the  simulation  
area and  stops  for duration $T_p$,  the node  then  randomly  and  uniformly  chooses   another  direction  to  travel.  As such,  the  nodes  remain  uniformly  distributed  with in the simulation area. Nevertheless, the transition lengths may deviate significantly from those in human walks. 

\vspace{0.2cm}
\subsubsection{Modified Random Direction Model~\cite{lin2013towards} }
Lin {\em et al.} updated the classical random direction  model by 
providing the transition length distributions directly as a Rayleigh distributed random variable in \cite{lin2013towards}. {\em They refer their model as modified random way point in \cite{lin2013towards}}. Through simulations, it was shown that, the modified model is close to Levy walk model which is close to human mobility patterns. The distribution of flight time was derived for constant velocity $V$ as follows~\cite{lin2013towards}:
\begin{equation}\label{tf1}
f_{T}(T)=2 \pi \lambda V^2 T e^{-\lambda \pi V^2 T^2}
\end{equation}
When $V$ is uniformly distributed in the range $[v_{\mathrm{min}} \quad v_{\mathrm{max}}]$, then $f_{T}(T)$ can be derived as follows~\cite{lin2013towards}:
\begin{equation}\label{tf2}
f_{T}(T)=\frac{g(v_{\mathrm{min}})- g(v_{\mathrm{max}}) }{t(v_{\mathrm{max}}-v_{\mathrm{min}})}
\end{equation}
where $g(x)=x e^{-\lambda \pi t^2 x^2} +\frac{\mathcal{Q} (t x\sqrt{2 \pi \lambda })}{\lambda t}$ in
which $\mathcal{Q}(\cdot)$ is the complimentary CDF (CCDF) of a Gaussian random variable. As in the classical RWP mobility model, the users can have a random pause time $T_p$ at each way point.


\vspace{0.2cm} 
\subsubsection{Synthetic Truncated Levy Walk Model~\cite{levy}} This model is constructed from real mobility trajectories. In the model, the transitions lengths and the pause time have an inverse power law distribution (Levy distribution as shown in Table~I). Flight times and lengths are found to be highly correlated, therefore \cite{levy} modeled the relation as $T= k L^{1-\rho}$. When $\rho =0$, $T$ becomes proportional to $L$ thus modeling constant velocity
movement; whereas, when $\rho=1$ flight time $T$ becomes a
constant and flight velocity is  proportional to $L$. 
Although Levy walk model is constructed from real mobility trajectories, human walks are not Levy walks since human walks  have complex temporal and spatial correlations and its nature has not yet been fully understood. 

\subsection{Memory-Based Mobility Models}
Due to the memoryless nature of the aforementioned mobility models, it is  difficult to capture the temporal dependency.  For example, the  current  
velocity or direction of a mobile node  may  depend  on the previous velocity and direction. As such, the velocities at different time epochs may be correlated.
In the following, few mobility models are discussed that are somewhat close to human mobility patterns and/or consider spatial and temporal correlations. 

\vspace{0.2cm}
\subsubsection{Smooth Random Mobility Model~\cite{bettstetterrwp}}

As has been noted before, the speed and direction of a node should vary incrementally and smoothly rather than randomly. In this context, an extension of random walk model, referred to as the smooth random mobility model, was  proposed in \cite{bettstetterrwp}. In this model, the PDF of velocity in the range $[0 \quad V_{\mathrm{m}}]$ can be explained by noting that the  preferred  speed  values of mobile node has  a  higher  probability, whereas a uniform velocity distribution is considered  on the remaining interval.
For  example,  if  the  node  has  the  preferred  speed set $\{ 0.25 V_{\mathrm{m}} \quad 0.5 V_{\mathrm{m}}\}$, then $f_V(v)$ can be derived as follows:
\begin{equation}\label{v}
f_V(v)=
\begin{cases}
\mathrm{Pr}(v= 0.5 V_{\mathrm{m}})\delta(v- 0.5 V_{\mathrm{m}}) & v= 0.5 V_{\mathrm{m}}\\
\mathrm{Pr}(v= 0.25 V_{\mathrm{m}})\delta(v- 0.25 V_{\mathrm{m}}) & v= 0.25 V_{\mathrm{m}}\\
\frac{1-\mathrm{Pr}(v= 0.25 V_{\mathrm{m}})-\mathrm{Pr}(v= 0.5 V_{\mathrm{m}})}{V_m} & 0 < V_m <1
\end{cases}
\end{equation}
where $\mathrm{Pr}(v= 0.5 V_{\mathrm{m}})+\mathrm{Pr}(v= 0.25 V_{\mathrm{m}}) <1$. In particular,  at each way point, the new speeds (or directions) are chosen from a weighted distribution of preferred speeds. 
The  frequency of speed change is a Poisson process, i.e., in
an  event  of  speed  change,  a  new  target speed 
is chosen according to \eqref{v}.  Then,  the  speed  changes   incrementally from the  current  speed  to  the  new   speed  by acceleration speed or deceleration speed  $a(t)$ whose PDF is uniformly distributed. 

\vspace{0.2cm}
\subsubsection {Gauss Markov Model~\cite{rwp0}} 
This model was introduced in \cite{rwp0} as an improvement over the Smooth Random mobility model \cite{bettstetterrwp}. In this model, a node's velocity at any time slot is a function  of its previous velocity, and thus, a temporal dependency exists.  The  degree  of  temporal dependency is determined by the parameter $\alpha$ which shows the memory level and randomness  of  the 
Gauss-Markov  process. When $\alpha=0$, the Gauss Markov process has strong memory and the velocity is determined by the fixed drift velocity and the Gaussian random variable. When $\alpha=1$, the Gauss Markov process becomes equivalent to fluid flow model where the velocity of the mobile node becomes equivalent to the previous velocity. For $0 \le \alpha \le 1$, the current velocity depends on both previous velocity and the Gaussian random variable. The  degree  of  randomness  is  adjusted  by $\alpha$. As $\alpha$ increases, the current velocity gets more influenced  by the previous velocity; otherwise, the current velocity becomes more  biased with the Gaussian random variable. 
At given intervals of time, the new speed and direction are calculated which is followed until the next time epoch. The dynamics of the model are greatly influenced by  $\alpha$ and the mean and standard deviation chosen for the Gaussian random variable.  Near the boundaries, the direction of movement is forced to flip 180 degree.

\vspace{0.2cm}
\subsubsection{Markovian Way Point Model~\cite{markovian}} This model is a generalized mobility model where the node moves along a straight line segment
from the current way point to the next way point (chosen from a discrete time Markov process). The way points constitute a Markov chain with finite transition probabilities. As such, the way points enable us to create popular routes a user walks or drives through, i.e., the consecutive legs may be correlated. The  velocity distribution can be designed based on the traveled distance, i.e., the longer transitions with vehicular speeds and shorter transitions with speeds
typical to a pedestrian user \cite{alparslan2007two}. The node can have a random pause time at each way-point as well as during the transitions.


\subsection{Lessons Learned}
Having  examined  the aforementioned  mobility  models,  we  observe  that  a trade-off exists between the analytical tractability and closeness to realistic  mobility   patterns.  Moreover, among the memoryless mobility models, the modified random direction and Levy flight models are the most recommended models due to their analytical tractability and closeness to human mobility patterns.
Note that the remaining memoryless models demonstrate a slow convergence towards the stationary distribution \cite{rwstationary}. Intuitively, once the mobile node chooses a far away destination with a slow speed, the node will get stuck in finishing this trip (due to low speed and far distance). Even with the increase in simulation runs, on average, more and more nodes can get stuck with long journeys and low speeds. A such,  the  average  nodal  speed  
keeps decreasing over time.   Interested readers can refer to \cite{rwpharmful,le2005stationary} for more details. 
A summary of the memoryless models is provided in Table~I. 

While the memoryless individual models are relatively tractable for performance analysis, they may not precisely capture the realistic movement patterns of wireless nodes such as high-speed trains, different variety of UAVs and drones, air crafts, vehicles, and human pedestrians. It is therefore important to incorporate  temporal correlation along with obstacle restrictions in  the mobility models and then utilize them for 5G/B5G wireless network applications. Finally, individual mobility models need to be generalized for group mobility models that are more suitable for high-speed vehicular applications. 

Later in the article, we will describe how memoryless individual mobility models are useful in the analysis of  wireless cellular networks with hexagonal grids, square grids, and single/multi-tier networks with BSs following a homogeneous PPP. For example,  \cite{lin2013towards} considered a random way point mobility model for a node and then conducted a mobility-aware performance analysis while \cite{bao2015stochastic} considered an arbitrary mobility pattern for handoff analysis.

\section{Current State-of-The-Art: Mobility and Handoff Analysis}

\subsection{Simulation-Based Studies}
To date, a plethora of research studies analyzed the mobility-based handoff performance in cellular networks via either computer simulations or numerical optimization algorithms.  For instance, \cite{ismail8} investigated the handoff performance of LTE networks. Self-organizing handoff management techniques were proposed in \cite{ismail9} to autonomously configure the mobility management parameters. In heterogeneous networks, handoff parameters (e.g., time-to-trigger (TTT), hysteresis threshold\footnote{It corresponds to the traffic load of BSs.}, etc.) were optimized to achieve seamless mobility of users in \cite{ismail15}. \cite{zorzi26} presented  multiple  vertical  handoff  decision  algorithms  for heterogeneous  networks, while 
\cite{zorzi27}  investigated  the   handoff  management
in multi-tier networks by proposing a theoretical model  to  characterize  the  performance  of  a  mobile  user  in heterogeneous  networks. 
In \cite{ismail18}, the significance of inter-cell interference coordination was shown to improve the handoff performance for both low and high speed users. Mobility state estimation was performed in \cite{ismail19} to
estimate the velocity of users and managing their associations accordingly, thereby enhancing the handoff performance. In \cite{ismail20}, mobility performance was analyzed with and without inter-site carrier aggregation for macrocells and picocells deployed on a different carrier frequencies.

The delays caused due to handoffs \cite{adve2015handoff,adve6} in multi-tier cellular networks were studied in various research works \cite{adve1,adve8,adve10}, where macrocells are overlaid by microcells~\cite{adve11,adve12}. The classic handoff algorithms typically assign users to the tier with maximum received signal power \cite{adve9}. In some research works, the users are classified as slow or fast according to a sojourn time threshold, and based on this,  the users will be assigned to a specific tier~\cite{adve13}. Some velocity-based handoff algorithms utilize the  estimated velocity and the direction of the user to perform handoff~\cite{adve8}. Nevertheless, to avoid the unnecessary ping-pong effect, some handoff schemes classify a given user as fast if it remains connected to the upper tier regardless of any changes in its speed~\cite{adve16}. Another alternative is to introduce a dwell-time threshold to take into account the history of the user (in terms of speed) before any handoff decision \cite{adve9}. Due to unnecessary or frequent handoffs between the tiers or among the BSs within one tier, the handoff schemes should be designed to reduce the handoff delay.  Other handoff criteria include degradation of the desired signal strengths, available resources/bandwidth at different BSs,    and  velocity  of mobile users. A detailed overview of such studies can be found in \cite{survey01,survey2, classic, comments,classicsurvey2}. 

\subsection{Theoretical Studies}
Despite aforementioned research on mobility management, there are limited research works that provide tractable mobility models and characterize network or users' performance  mathematically with mobility. Performance characterization of a cellular network  enables network operators to qualitatively assess and optimize the  performance beforehand without expensive field trials or time-consuming simulations. Approximations may also be developed that can potentially simplify the optimization procedures and may provide closed-form optimal solutions for important network performance metrics. 
In this context, some of the classical mobility-aware performance models employed queuing, i.e., the cells are generally modeled as queues, active users are modeled as units in the queues, and handoffs correspond to unit transfers among queues. For example,  \cite{bao15} studied the single-cell set-up using an
M/G/$\infty$ queue and  \cite{bao16} proposed a two-queue model considering a wireless local-area network (WLAN) cell overlaying one 3G cell. For multi-cell networks, queuing was employed in \cite{bao17, bao18,bao19,bao20, bao21, bao22}. {\em The impact of cellular network geometry was however not considered}.

A few of the research works incorporated spatial features of cellular networks by analyzing single-tier \cite{bao23} and multi-tier networks \cite{bao26} with  BSs deployed on regular grids. In \cite{bao23},  a mathematical formulation was developed for the random movement of a user in a single-tier hexagonal cellular network.   The proposed model characterized the  distribution  of  the  cell  residence time (sojourn time),  the  channel  holding time (defined as the time during which a  call occupies a channel in the given cell),  and  the  average  number  of  handoffs  per  user.  Results showed  that  the  generalized  gamma  distribution  is  adequate  to describe the cell residence time distribution. It was also shown that the negative exponential distribution  is  a  good  approximation  for  the  channel  holding time distribution. In this work, no specific user mobility model was considered. \cite{bao26} studied a two-tier cellular network including one circular macrocell and a predetermined
number of circular microcells. A mobility model was proposed to calculate the user transition probabilities from different cells assuming that the sojourn time distribution is given. 

\subsection{Lesson Learned}
The aforementioned studies considered non-random spatial tessellation of cellular BSs and therefore may not be suitable for dense, heterogeneous, and random 5G/B5G networks.  Also, performance metrics such as mobility-aware SINR coverage and mobility-aware throughput were not considered.

\section{Mobility-Aware Performance Analysis of Random Cellular Networks}

Recently,  in a handful of studies, mobility has been considered for random cellular networks.   The approaches for mobility-aware analysis of random cellular networks  can be classified as follows:
\begin{itemize}

\item {\em Approach~1 (Trajectory-Based Approach)~\cite{lin2013towards,bao2015stochastic}}:  This approach assumes that the handoff event occurs when the moving user  travels across cell boundaries of different BSs along  its  trajectory (which can be defined according to a mobility pattern described in Section~II).  Then  the  analysis  of  handoff  rate  involves  the  evaluation  of  the  number  of  intersections between  the  user  trajectory  and  the  set  of  cell  boundaries.  
This approach requires the derivation of the  statistical  distribution  of  cell boundaries which is relatively complex. This approach leads to handoff rate and sojourn time evaluation.

\item {\em Approach~2 (Association-Based Approach)~\cite{adve2015handoff,shin2017equivalent}}:  This approach assumes that the handoff event occurs when there is a neighboring BS that provides a stronger signal quality than the serving BS.  That is, this approach leads to the evaluation of the probability of handoff during one movement period using the user association criterion and association probability. 

\end{itemize}
In the following, we review some of the pioneering research works where the aforementioned mobility-aware performance analysis techniques were considered.

\subsection{Trajectory-Based Approach}
 \cite{lin2013towards} utilized the modified random direction model to mimic the movement of users  in a single-tier cellular network with BSs deployed regularly as  hexagonal lattices and randomly following a homogeneous PPP.  Analytical expressions were derived for the handoff rate (i.e., the average number of cells a mobile user traverses to the average transition time (including the pause time)) and sojourn time. The transition length  was considered to be i.i.d. Rayleigh distributed which is contrary to  classical RWP model where the  transition lengths are not i.i.d. and the random way points are i.i.d. 

The general approach for mobility-aware performance analysis in  this approach can be summarized as follows:

\vspace{2mm}
{\em\underline {Step~1: Characterize  Mobility-Related Parameters}}
\vspace{2mm}

\begin{itemize}

\item Derive the expected transition length $\mathbb{E}[L]$.
\item Derive the expected velocity $\mathbb{E}[V]$.
\item Derive the distribution and expected value of the flight time {$T=L/V$}.
\item Derive the mean of the pause time $S$.
\item Derive the distribution of the target way point ${\bf X_1}$ which is required for  the evaluation of handoff rate. Conditioned on the position of $X_1$, the number of handoffs equals the number of intersections of the segment $[X_0,X_1]$ and the boundary of the Poisson-Voronoi tessellation. The expected number of handoffs can therefore be derived by averaging over the distribution of $X_1$ and the distribution of Poisson-Voronoi tessellation.
\item Derive the spatial node distribution $f(r,\theta)$ which is defined as the probability that the  node resides in some measurable set $\mathcal{A}$ while transitioning from ${\bf X_0}$ to ${\bf X_1}$. The distribution  is needed for sojourn time computation.
\end{itemize}

\vspace{2mm}
{\em \underline{Step~2:  Handoff Rate Using Buffon's Needle Problem}}
\vspace{2mm}

To derive the handoff rate for a user that moves with velocity $v$ in a time period of $\Delta t$ {in area $\mathcal{A}$}, we need to calculate the probability that this user crosses the cell boundaries. That is,  find the probability that a needle with length 
$\ell=v \Delta t$ crosses a line, where the lines are the cell boundaries. 
This  problem is referred to as the Buffon's needle problem. 
\begin{definition}[Buffon's Needle Problem \cite{schroeder1974buffon}]{
A needle of length $\ell$ is dropped onto a floor with equally spaced parallel lines. What is the probability that needle crosses a line?}
\end{definition}
{Using the generalized argument for Buffon's needle problem\footnote{{Buffon's needle problem was studied for both short (i.e., length of the needle is less than the distance between two lines) and long needles. In this article, since we are considering short time period of $\Delta t$, we need the result derived for short needle.}}, for a user that moves in direction $\theta$, the handoff rate (conditioned on the $\theta$ and velocity) can be obtained as \cite{lin2013towards}  
	\begin{IEEEeqnarray}{rCl}
		H= v |\sin{\theta}| \lim_{|\mathcal{A}|\to\infty} \frac{|\mathcal{B}_{\mathcal{A}}|}{|\mathcal{A}|}  \frac{\mathbb{E}[T]}{\mathbb{E}[T]+\mathbb{E}[S]}.
		\label{eq:handoff_rate_0}
	\end{IEEEeqnarray}
	In the modified random direction model,  $\theta$ is uniformly distributed on $[0,2\pi]$. Therefore, $\mathbb{E}\left[ |\sin{\theta}| \right]=2/\pi$ which yields\footnote{
		{\eqref{eq:handoff_rate_1} can be used for any mobility model where $\theta$ is uniformly  distributed.}} 
}
\begin{IEEEeqnarray}{rCl}
	H=\frac{2}{\pi} \mathbb{E}[V] \lim_{|\mathcal{A}|\to\infty} \frac{|\mathcal{B}_{\mathcal{A}}|}{|\mathcal{A}|} \frac{\mathbb{E}[T]}{\mathbb{E}[T]+\mathbb{E}[S]}
	\label{eq:handoff_rate_1}
\end{IEEEeqnarray}
where $T$ denotes the transition time and $S$ denotes the pause time, i.e., ${\mathbb{E}[T]}/\left({\mathbb{E}[T]+\mathbb{E}[S]}\right)$ is the expected proportion of transition time. $|\mathcal{A}|$ is the {size of} area $\mathcal{A}$ and $|\mathcal{B}_{\mathcal{A}}|$ is the length of cell boundaries {in $\mathcal{A}$}. Note that $\lim_{|\mathcal{A}|\to\infty} \frac{|\mathcal{B}_{\mathcal{A}}|}{|\mathcal{A}|}$ is the average length of the cell boundaries in a unit area which in \cite{bao2015stochastic} is called {\em length intensity of cell boundaries}, denoted by $\mu_{1}( \text{T}^{(1)} )$. Therefore,  
\begin{IEEEeqnarray}{rCl}
	H=\frac{2}{\pi} \mathbb{E}[V] \mu_{1}( \text{T}^{(1)} ) \frac{\mathbb{E}[T]}{\mathbb{E}[T]+\mathbb{E}[S]},
	\label{eq:handoff_rate_2}
\end{IEEEeqnarray}
where $\text{T}^{(1)}$ represents set of cell boundaries. The length intensity of cell boundaries can then be defined as follows.
\begin{definition}[Length Intensity of Cell Boundaries \cite{bao2015stochastic}]
	{ Length intensity of the cell boundaries, denoted by $\mu_{1}(\text{\rm T}^{(1)})$, is the average length of $\text{\rm T}^{(1)}$ in a unit square. }
\end{definition}
{\bf Remark:} 
The handoff rate for hexagonal lattice can be obtained by using the generalized argument of the Buffon's needle problem.  However, for single-tier PPP, the handoff rate was derived using an alternative approach in \cite{lin2013towards}. In \cite{bao2015stochastic}, the authors provided a general approach to model the handoff rate by deriving the length intensity of cell boundaries considering  single-tier as well as multi-tier PPP.
The  approach is applicable for random multi-tier BSs, arbitrary user movement trajectory, and flexible user-BS association. A connection between the network performance metrics such as coverage probability and network ergodic capacity  was not established in \cite{lin2013towards,bao2015stochastic}. Nevertheless, \cite{hesham2016split,hesham2017velocity,heshamletter} incorporated the handoff rate calculations from \cite{bao2015stochastic} to derive mobility-aware throughput.

\vspace{2mm}
{\em \underline{Step~3:  Length Intensity  of Cell Boundaries}}
\vspace{2mm}

Evidently, to derive the handoff rate, we need to calculate the length intensity of cell boundaries $\mu_{1}( \text{T}^{(1)} )$. In the following,  the steps to derive $\mu_{1}( \text{T}^{(1)} )$ are discussed \cite{bao2015stochastic}. 
\begin{itemize}
	\item First, we need to derive area intensity of $\Delta d$-extended cell boundaries, $\mu_{2}(\text{\rm T}^{(2)}(\Delta d))$, which is the average area of $\text{T}^{(2)}(\Delta d)$ in a unit square, where $\text{T}^{(2)}(\Delta d)$ denotes the $\Delta d$-extended cell boundaries and is defined as the set of points that are located within (perpendicular) distance $\Delta d$ from $\text{T}^{(1)}$, i.e., 
		\begin{IEEEeqnarray}{rCl}
			\text{T}^{(2)}(\Delta d)= \left\{ x\in\mathbb{R}^2 :\exists y\in \text{T}^{(1)}, 
			\hspace{2pt} \text{s.t.} \hspace{2pt} |x-y| < \Delta d \right\}. \nonumber \\
			\label{eq:delta_d_extended}
		\end{IEEEeqnarray}
		$\mu_{2}(\text{\rm T}^{(2)}(\Delta d))$ can be interpreted as the probability that a randomly located point in a unit square is in $\text{\rm T}^{(2)}(\Delta d)$. In a cellular network, BSs locations are fixed and users are randomly distributed. However, from the perspective of each user, the BS realization is different. Instead of considering a fixed realization of the BSs and a randomly located user, we assume that the user is always located at the origin and the BSs are randomly distributed. Therefore, 
		$\mu_{2}(\text{\rm T}^{(2)}(\Delta d))$ is the probability that the origin ($\mathbf{0}$) is in $\text{\rm T}^{(2)}(\Delta d)$, i.e.,  
		\begin{IEEEeqnarray}{rCl}
			\mu_{2}(\text{\rm T}^{(2)}(\Delta d))=\mathbb{P}\left( \mathbf{0} \in \text{\rm T}^{(2)}(\Delta d) \right).
			\label{eq:area_intensity}
		\end{IEEEeqnarray}
		Cell boundaries are given by the BS locations. Thus, the event that the origin is in $\text{\rm T}^{(2)}(\Delta d)$ is equal to the event that the BSs are located such that the set of cell boundaries 
		$\text{\rm T}^{(1)}$ is within distance $\Delta d$ from the origin.
	\item Then the length intensity is obtained by
	\begin{IEEEeqnarray}{rCl}
		\mu_{1}(\text{\rm T}^{(1)})	= \lim_{\Delta d \to 0} \frac{ \mu_{2}(\text{\rm T}^{(2)}(\Delta d)) }{2 \Delta d}.
		\label{eq:length_intensity}
	\end{IEEEeqnarray}
\end{itemize}


\begin{figure}[t]
	\centering
	\includegraphics[width=0.5\textwidth]{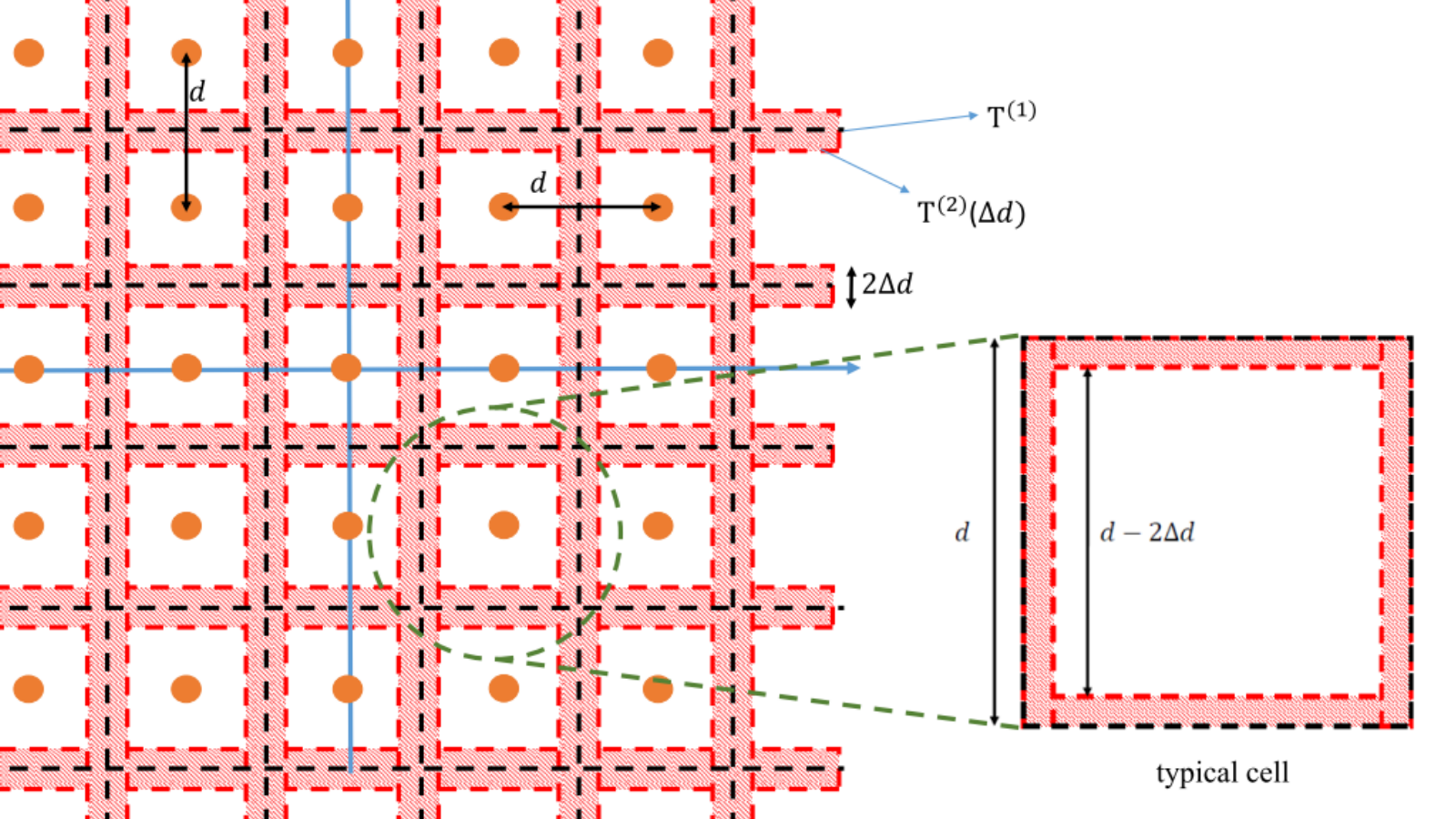}
	\caption{
		Square lattice with spacing $d$. The orange dots represent the BS locations, the black dashed lines represent the cell boundaries $\text{\rm T}^{(1)}$, and the shaded red area represents $\text{\rm T}^{(2)}(\Delta d)$. }
	\label{fig:square_lattice_1}	
\end{figure}

\begin{figure*}[t]
\begin{center}
\includegraphics[width=\textwidth]{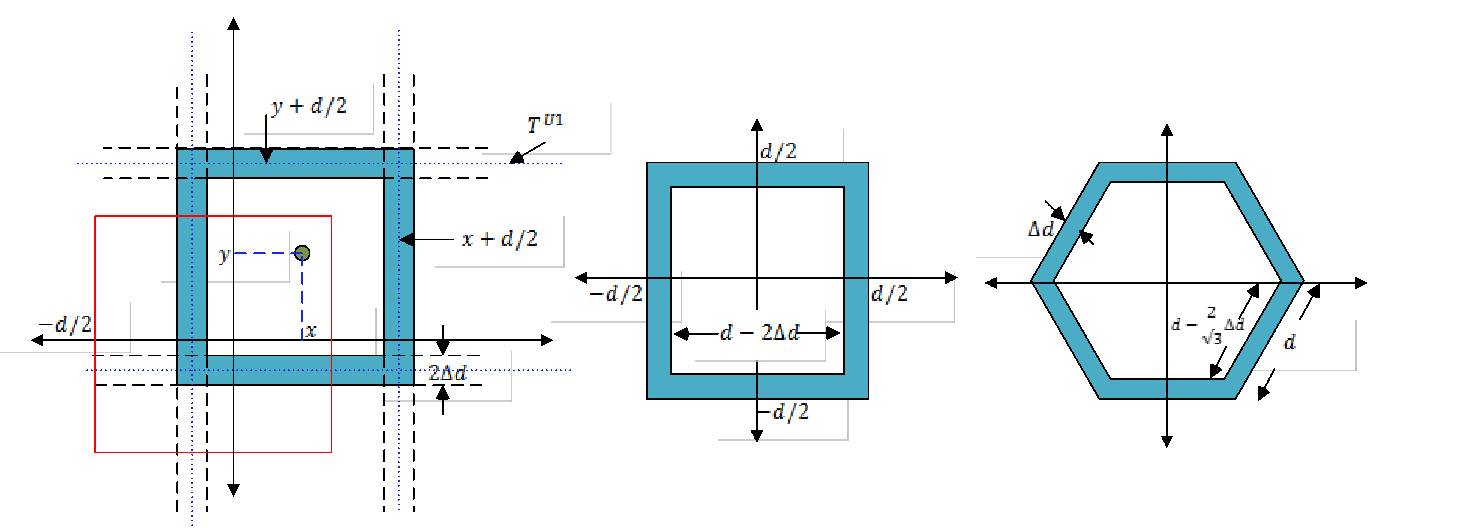}
\end{center}
\caption{Graphical illustration for the length intensity $\text{\rm T}^{(1)}$ and area intensity $\text{\rm T}^{(2)}(\Delta d)$ of the cell boundaries in square and hexagonal spatial cellular networks.}
\label{fig:3}
\end{figure*}

\subsection{Case Studies and a Recommended Methodology}

{Here we provide three case studies in which we derive handoff rates for square lattices, hexagonal grids, and three dimensional (3-D) single-tier PPP networks (e.g., for drone applications)} by combining Buffon's needle approach from \cite{lin2013towards} with length intensity of cell boundaries from \cite{bao2015stochastic} as it can be extended to multi-tier cellular networks. Also, we show that this approach can be easily applied to both random and non-random cellular networks.

\subsubsection{\bf Case Study~1 (Square Lattice)} {Assume that the BS locations are modeled by a square lattice with spacing $d$\footnote{In 2-D space, square lattice  with spacing $d$ is defined as $\mathbb{L}_d = \left\{ d \mathbb{Z}^2 \right\}$~\cite{haenggi2012stochastic}.} as shown in \figref{fig:square_lattice_1}. The set of cell boundaries $\text{\rm T}^{(1)}$ and $\Delta d$-extended cell boundaries $\text{\rm T}^{(2)}(\Delta d)$ are also illustrated in \figref{fig:square_lattice_1}. Assume that a user is randomly located in the typical cell and we call this user the typical user. Then $\mu_{2}(\text{\rm T}^{(2)}(\Delta d))$ is the probability that the typical user is in the shaded area, i.e.,
	\begin{IEEEeqnarray}{rCl}
		\mu_{2}(\text{\rm T}^{(2)}(\Delta d))=1-\frac{( d-2\Delta d )^2}{d^2}=\frac{4 \Delta d}{d}+ O(\Delta d^2).\qquad
\end{IEEEeqnarray}}

{As we have mentioned earlier, we can assume that the typical user is always at the origin and the BSs are randomly distributed,} i.e., one BS (the reference BS) is located uniformly in the square $[-d/2,d/2]^2$ and the other BSs are located with spacing $d$. Let us denote the location of the reference BS by $(x,y)$. Therefore, the BSs in the neighboring cells are located at $(x,y+d)$, $(x,y-d)$, $(x+d,y)$, $(x-d,y)$ and the  cell boundaries are $y+d/2$, $y-d/2$, $x+d/2$, and $x-d/2$. These boundaries are shown in \figref{fig:3}{(a)}. {The blue area in \figref{fig:3}(a) shows $\Delta d$-extended region of the cell boundaries $\text{\rm T}^{(2)}(\Delta d)$.}

{According to \eqref{eq:area_intensity}, $\mu_{2}(\text{\rm T}^{(2)}(\Delta d))$ is the probability that $\mathbf{0}\in\text{\rm T}^{(2)}(\Delta d)$. $\mathbf{0}\in\text{\rm T}^{(2)}(\Delta d)$ means that the origin is within distance $\Delta d$ from one of the cell boundaries, i.e., $|y+d/2-\mathbf{0}|<\Delta d$, $|y-d/2-\mathbf{0}|<\Delta d$, $|x+d/2-\mathbf{0}|<\Delta d$, or $|x-d/2-\mathbf{0}|<\Delta d$. Simplifying these four inequalities and considering $-d/2<x,y<d/2$, yields $-d/2<y<-d/2+\Delta d$, $d/2-\Delta d<y<d/2$, $-d/2<x<-d/2+\Delta d$, or $d/2-\Delta d<x<d/2$. The shaded area in \figref{fig:3}(b) illustrates these four inequalities for $-d/2<x,y<d/2$. Consequently, $\mathbf{0}\in\text{\rm T}^{(2)}(\Delta d)$ is equal to the event that the reference BS located at $(x,y)$ falls within the shaded area in \figref{fig:3}(b). It is worth mentioning that the shaded area in \figref{fig:3}(b) is exactly the same as the shaded area in the typical cell in \figref{fig:square_lattice_1}.}

From \eqref{eq:length_intensity}, the length intensity is $\mu_{1}(\text{\rm T}^{(1)})=2/d$, and using \eqref{eq:handoff_rate_2} we can derive the handoff rate as
\begin{IEEEeqnarray}{rCl}
	H=                  \frac{4}{\pi} \frac{\mathbb{E}[V]}{d} \frac{\mathbb{E}[T]}{\mathbb{E}[T]+\mathbb{E}[S]} 
	\stackrel{(a)}{=} \frac{4}{\pi }      \mathbb{E}[V] \sqrt{\lambda}  \frac{\mathbb{E}[T]}{\mathbb{E}[T]+\mathbb{E}[S]}, \qquad
	\label{eq:hanover_square}
\end{IEEEeqnarray}
where $\lambda$ in (a) denotes the BS density (average number of BSs in a unit area), which can be obtained by $\lambda={1}/{d^2}$.

\subsubsection{\bf Case Study~2 (Hexagonal Grid)} For the hexagonal lattice with side length of $d$, i.e., 
$({3\sqrt{3}}/{2})d^2=1/\lambda$, the area intensity of the $\Delta d$-extended cell boundaries is equal to the probability that the reference BS (randomly distributed in the hexagon) falls in the shaded area in \figref{fig:3}(c), i.e.,
\begin{IEEEeqnarray}{rCl}
	\mu_{2}(\text{\rm T}^{(2)}(\Delta d))=1-\frac{ \frac{3\sqrt{3}}{2}( d-\frac{2}{\sqrt{3}}\Delta d )^2 }{ \frac{3\sqrt{3}}{2}d^2 }
	= \frac{4 \Delta d}{\sqrt{3}d}+ O(\Delta d^2). \nonumber
\end{IEEEeqnarray}
The handoff rate  can then be obtained as follows~\cite{lin2013towards}:
\begin{IEEEeqnarray}{rCl}
	H=                  \frac{4}{\sqrt{3} \pi} \frac{\mathbb{E}[V]}{d} \frac{\mathbb{E}[T]}{\mathbb{E}[T]+\mathbb{E}[S]}.
	\label{eq:hanover_hexagonal} 
\end{IEEEeqnarray}

\subsubsection{{\bf Case Study~3 (3-D Single-tier PPP)}} Consider a single-tier cellular network, where the BS locations are modeled by a homogeneous PPP of intensity $\lambda$. We assume that all the BS antennas are installed at the same height from the ground. We want to derive the handoff rate for a mobile drone which is associated to these BSs. The location of the drone at time $t+1$, $t\in\mathbb{N}$, is denoted by $\text{x}(t+1)$ and assume that $\text{x}(t+1)$ is uniformly distributed within the sphere of radius $(4/3)\bar{v}$ (the term (4/3) is used to normalize the average velocity) centered at $\text{x}(t)$, i.e., for $t\in\mathbb{N}$,
	\begin{equation}
	\text{x}(t+1)=\text{x}(t)+
	\begin{bmatrix}
	v(t)\sin(\varphi(t))\cos(\theta(t)) \\
	v(t)\sin(\varphi(t))\sin(\theta(t)) \\
	v(t)\cos(\varphi(t)) 
	\end{bmatrix}
	\end{equation}
	where $v(t)$ is the velocity of the drone during $t$-th movement period; $\varphi(t)$ and $\theta(t)$ also determine the moving direction during the $t$-th movement period. Since $\text{x}(t+1)$ is uniformly distributed within the 3-D ball of radius $(4/3)\bar{v}$ centered at $\text{x}(t)$,
	$v(t)$ is i.i.d. over $t$ and its distribution is given by 
	$f_v(v)=81 v^2/(64 \bar{v}^3)$, where $v\in[0,(4/3)\bar{v}]$; therefore, $\mathbb{E}[v]=\bar{v}$. $\varphi(t)$ is also i.i.d. over time and $f_{\varphi}(\varphi)=\frac{1}{2}\sin\varphi$, where $\varphi\in[0,\pi]$. Moreover, at time $t$, $\theta(t)$ is uniformly distributed in $[0,2\pi]$. We assume that all the BSs transmit with the same power; therefore, we consider the nearest BS association. Since the BS antennas are installed at the same height from the ground, the set of cell boundaries consists of planes that are orthogonal to the $\mathbb{R}^2$ plane so that the set of cell boundaries can be fully defined by its intersection with $\mathbb{R}^2$ plane (which is the same as the set of cell boundaries for a ground user). Therefore, we can study the handoff rate of the drone by projecting its trajectory onto $\mathbb{R}^2$ plane, i.e., the handoff rate of the drone is the same as the handoff rate of a ground user that moves with velocity $v(t)\sin(\varphi(t))$ in the direction $\theta(t)$ in time interval $[t,t+1)$. If we assume no pause time, from \eqref{eq:handoff_rate_0}, we have 
	\begin{IEEEeqnarray}{rCl}
		H = \mathbb{E}[v]  \mathbb{E}[\sin\varphi] \mathbb{E}\left[|\sin{\theta}|\right] \mu_{1}(\text{T}^{(1)})  
		&\stackrel{(a)}{=}& \bar{v} \sqrt{\lambda} 
	\end{IEEEeqnarray}
	where (a) is obtained from (i) $\mu_{1}(\text{T}^{(1)})=2 \sqrt{\lambda}$ for the single-tier PPP network \cite{bao2015stochastic}, (ii) $\mathbb{E}[\sin\varphi]=\pi/4$ for the mentioned distribution, and (iii) $\mathbb{E}\left[|\sin{\theta}|\right]=2/\pi$ since $\theta$ is uniformly distributed in $[0,2\pi]$.

\subsubsection{Results}
{\figref{fig:Hanover_1} compares the handoff rates observed in various cellular network settings such as  square lattice \eqref{eq:hanover_square}, hexagonal grid lattice \eqref{eq:hanover_hexagonal}, and single-tier PPP \cite{lin2013towards,bao2015stochastic} for various user velocities when $\lambda=0.0004$. The handoff rate for multi-tier PPP \cite{bao2015stochastic} is also shown in this figure (for two-tier networks). {Based on the analysis, the handoff rates for square lattice and single-tier PPP are the same since the length intensities of the cell boundaries are the same and equal to $2\sqrt{\lambda}$. This is also verified from the simulations in \figref{fig:Hanover_1}.} Moreover, the hexagonal lattice shows a slightly reduced handoff rate compared to the square lattice and single-tier PPP.} 


\subsection{Association-Based Approach}
In \cite{adve2015handoff}, the authors considered  $K$-tier (orthogonal spectrum allocated to different tiers) PPP network model for handoff and coverage analysis of a mobile user moving at speed $v$ from one point to another. 
The handoff probabilities provided in \cite{adve2015handoff} are not precise. In the following, first we will provide a brief overview of their methodology and then  provide the exact expression for handoff probability.

\begin{figure}[t]
	\centering
	\includegraphics[width=0.4\textwidth]{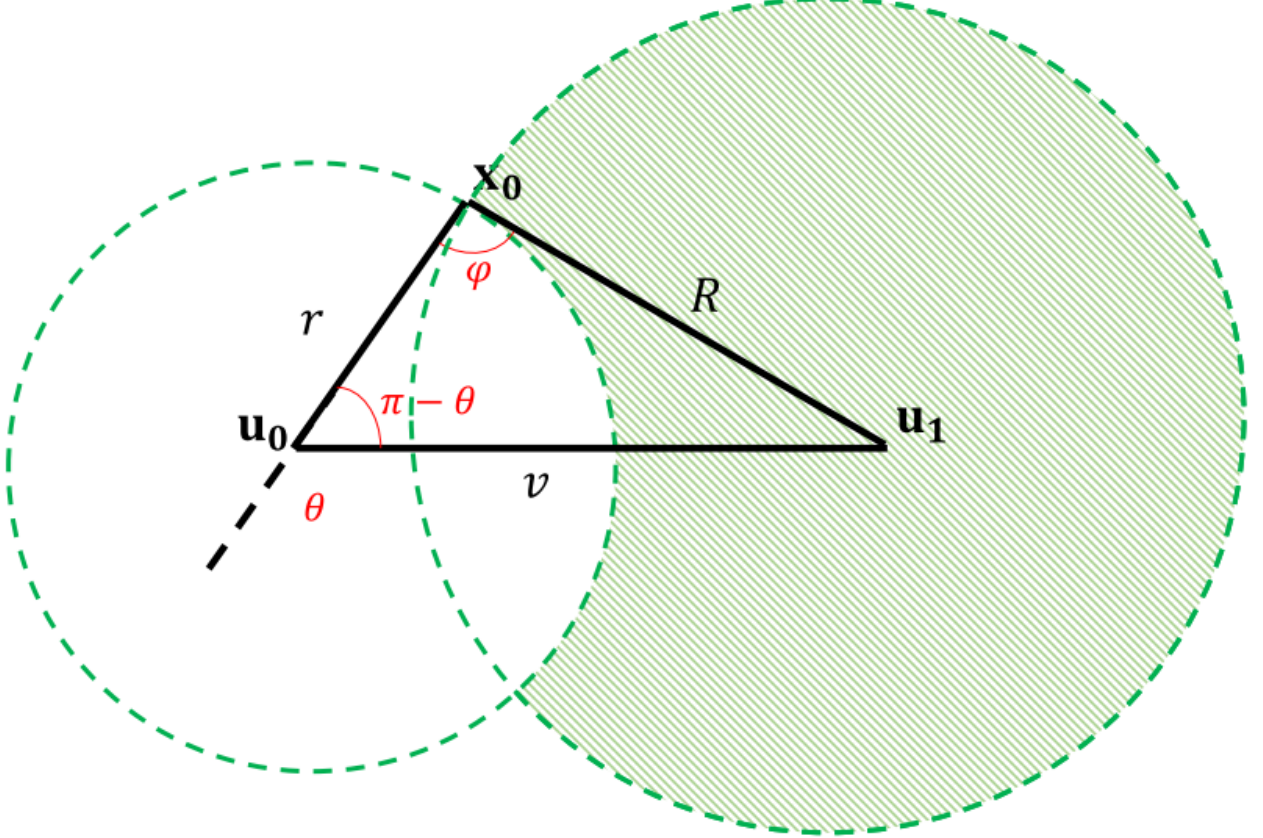}
	\caption{System model in \cite{adve2015handoff}.}
	\label{fig:geometry}	
\end{figure}

Consider a single-tier Poisson cellular network where BSs follow a homogeneous PPP $\Phi$ of intensity $\lambda$ and each user connects to its nearest BS. Assume a user located at $\text{\bf u}_0$ is connected to a BS located at $\text{\bf x}_0$ as shown in \figref{fig:geometry}. Let us denote the distance between the BS and the user by $r$, i.e., $r=\|\text{\bf u}_0-\text{\bf x}_0\|$. Since the user is connected to its nearest BS, there is no BS in the ball of radius $r$ centred at $\text{\bf u}_0$.
Now assume that this user moves with velocity $v$ at angle $\theta$ {(with respect to the direction of connection)} and its new location is $\text{\bf u}_1$. Using the law of cosines, $\text{\bf u}_1$ is at distance $R=\sqrt{r^2+v^2+2rv\cos(\theta)}$ from the BS located at $\text{\bf x}_0$. 

\vspace{0.2cm}
\subsubsection{Handoff Probability} 
Clearly, {handoff occurs when a different BS is closer to $\text{\bf u}_1$ than the BS located at $\text{\bf x}_0$, i.e., when there is at least one BS in the shaded green area in \figref{fig:geometry}}.
Therefore, given $r$ and $\theta$ the probability of handoff is 
\begin{IEEEeqnarray}{rCl}  
	\mathbb{P}(H \mid r,\theta) &=&\mathbb{P} 
	\left( \Phi\left( b\left(\text{\bf u}_1,R\right) \setminus b\left(\text{\bf u}_0,r\right) \right)>0 \mid r,\theta \right) 
	\nonumber \\
	&\stackrel{(a)}{=}& 
	1-\exp\left( -\lambda |b\left(\text{\bf u}_1,R\right) \setminus b\left(\text{\bf u}_0,r\right)| \right) \nonumber
\end{IEEEeqnarray}
where $b\left(\text{\bf u}_1,R\right)$ denotes the ball with radius $R$ centred at $\text{\bf u}_1$ and $b\left(\text{\bf u}_0,r\right)$ is excluded from $b\left(\text{\bf u}_1,R\right)$ since we know the BS located at $\text{\bf x}_0$ is the nearest BS to $\text{\bf u}_0$. Note that (a) is obtained from the void probability of the PPPs and $|b\left(\text{\bf u}_1,R\right) \setminus b\left(\text{\bf u}_0,r\right)|$ denotes the area of 
$b\left(\text{\bf u}_1,R\right) \setminus b\left(\text{\bf u}_0,r\right)$. Finally, the handoff probability is obtained by averaging over $r$ and $\theta$, {where $\theta$ is uniformly distributed between $0$ and $\pi$ (due to the symmetry), and $r$ is a Rayleigh random variable with mean $1/(2\sqrt{\lambda})$, i.e., $f_r(r)= 2 \lambda \pi r e^{- \lambda \pi r^2}$.
Following these steps, $\mathbb{P}(H)$ for single-tier PPP can be derived as follows~\cite{adve2015handoff}:
\begin{IEEEeqnarray}{rCl}
	\IEEEeqnarraymulticol{3}{l}{
	\mathbb{P}(H )= \frac{1}{\pi} 
	\int_{0}^{\pi} \int_{0}^{\infty} \mathbb{P}(H \mid r,\theta) 2 \lambda \pi r e^{- \lambda \pi r^2} {\rm d}r {\rm d}\theta}
	\label{eq:handoff_probability}
\end{IEEEeqnarray}
where $\mathbb{P}(H \mid r,\theta)$ is given as~\cite{adve2015handoff}:
\begin{IEEEeqnarray}{rCl}
	1-[e^{-\lambda \left( 
		R^2 \left[ \pi - \theta + \sin^{-1}\left( \frac{v\sin\theta}{R}\right) \right] -r^2(\pi-\theta)+rv\sin\theta
		\right)}].
	\label{eq:condition_handoff_probability}
\end{IEEEeqnarray}}

{\bf Correction:} For $\theta=\pi$, the above handoff probability calculation is precise when $v \leq r$. This can easily be understood by setting $\theta=\pi$ in \eqref{eq:condition_handoff_probability} which gives $\mathbb{P}(H \mid r,\theta)=0$. For $\theta=\pi$, $\mathbb{P}(H \mid r,\theta)=0$ is only true when $v\le r$. Therefore, for $v>r$ we must use the following equation:
\begin{IEEEeqnarray}{rCl}
	\mathbb{P}(H \mid r,\theta)=1-e^{-\lambda \left( 
		R^2 \left[ 2 \pi - \theta - \sin^{-1}\left( \frac{v\sin\theta}{R}\right) \right] -r^2(\pi-\theta)+rv\sin\theta
		\right)} \nonumber \\
	\label{eq:condition_handoff_probability_modified}
\end{IEEEeqnarray}
which is not zero for $\theta=\pi$. 
Since $\theta$ is between 0 and $\pi$, 
$\varphi=\sin^{-1}\left( \frac{v\sin(\theta)}{R}\right)$ in \eqref{eq:condition_handoff_probability} is always between $0$ and $\pi/2$. 
$\varphi=\angle {\text{\bf u}_0}{\text{\bf x}_0}{\text{\bf u}_1}$, i.e., $\varphi$ is the angle of the vertex ${\text{\bf x}_0}$ in $\triangle {\text{\bf u}_0}{\text{\bf x}_0}{\text{\bf u}_1}$ as illustrated in \figref{fig:geometry}.
$\varphi$ is between $0$ and $\pi/2$ when $\theta$ is between 0 and $\pi/2$ or $v\cos(\pi-\theta)\le r$. When $\theta$ is between $\pi/2$ and $\pi$ and $v\cos(\pi-\theta)> r$, $\sin^{-1}\left( \frac{v\sin(\theta)}{R}\right)$  in \eqref{eq:condition_handoff_probability} must be replace by $\pi-\sin^{-1}\left( \frac{v\sin(\theta)}{R}\right)$ since $\varphi$ is between $\pi/2$ and $\pi$. Therefore, the exact expression for the handoff probability is given by  \eqref{eq:handoff_probability_modified}, where $A=\left[ \pi - \theta + \sin^{-1}\left( \frac{v\sin(\theta)}{R}\right) \right]$.
\begin{figure*}
\begin{align}
		&\mathbb{P}(H )=  
	 1 - {2 \lambda} \Bigg[ \int\limits_{0}^{\pi/2} \int\limits_{0}^{\infty} r e^{-\lambda \left( R^2 A +r^2\theta+rv\sin(\theta)
		\right)}{\rm d}r {\rm d}\theta
	+
 \int\limits_{\pi/2}^{\pi} \int\limits_{0}^{v \cos(\pi - \theta)} r e^{-\lambda \left( 
		R^2 \left[ 2\pi - \theta - \sin^{-1}\left( \frac{v\sin(\theta)}{R}\right) \right] +r^2\theta+rv\sin(\theta)
		\right)}{\rm d}r {\rm d}\theta 
 \nonumber \\
&+ \int\limits_{\pi/2}^{\pi} \int\limits_{v \cos(\pi - \theta)}^{\infty}  r e^{-\lambda \left( 
		R^2 A +r^2\theta+rv\sin(\theta)
		\right)}{\rm d}r {\rm d}\theta\Bigg]. 
	\label{eq:handoff_probability_modified}
\end{align}
\hrule
\end{figure*}

\vspace{0.2cm}
\subsubsection{Mobility-Aware Coverage Probability} The coverage probability (or transmission success) of a typical user, conditioned on $r$ and $\theta$, can then be defined as follows:
\begin{align}\label{1}
\mathbb{C}_k=&\mathbb{P}(\gamma_k \geq \tau_k, \bar H_k) + (1-\beta)\mathbb{P}(\gamma_k \geq \tau_k, H_k)
\end{align}
where $\beta$ is the probability of handoff failure due to dropped connections even though the user is in coverage from SIR point of view. The coefficient $\beta$ represents the system sensitivity to handoffs and its value depends on the radio access technology, the mobility protocol, and the speed of the link. When $\beta=1$, every handoff leads to an outage irrespective of SIR coverage, whereas for $\beta=0$, the coverage of a user is not sensitive to handoffs  and becomes equal to the traditional coverage probability of a user.

For multi-tier networks, the coverage probability of a user who is initially associated to tier $k$ was defined as follows:
\begin{multline}\label{multi}
\mathbb{C}_k=\sum_{k=1}^{K} (1-\beta) \mathbb{P}(\gamma_k \geq \tau_k, n=k) \\+ \beta \mathbb{P}(\gamma_k \geq \tau_k, n=k, \bar H_k)
\end{multline}
which can be equivalent written as follows:
\begin{multline}\label{multi1}
\mathbb{C}_k=\sum_{k=1}^{K} \mathbb{P}(\gamma_k \geq \tau_k, n=k, \bar H_k) + \\(1-\beta) \mathbb{P}(\gamma_k \geq \tau_k, n=k, H_k).
\end{multline}
{\bf Correction:} The formulation in \eqref{multi} or \eqref{multi1} considers the SIR of the user with tier $k$ even after handoff occurs which cannot precisely capture the inter-tier handoff as the user is now connected to the $j$-th tier. For multi-tier networks, the coverage probability of a user initially associated to tier $k$ should be defined as follows: 
\begin{multline}
\mathbb{C}_k=(1-\beta) \sum_{j=1}^{K} \mathbb{P}(\gamma_j \geq \tau_j, n=k, H_{k,j}) \\+ \mathbb{P}(\gamma_k \geq \tau_k, n=k, \bar H_k)
\label{correction}
\end{multline}
where $\mathbb{P}(n=k)$ is given by the association probability to tier $k$.
Another limitation is that the closest BS to the user after handoff is always identified as the new serving BS, which  may  not be  true  in  the downlink of multi-tier cellular networks (due to distinct transmission powers and received signal powers).  As such, the results are limited to  nearest BS association (which is valid in case of single-tier downlink networks). These results precisely elaborate horizontal handoff probability while ignoring the essence of multi-tier networks. The biased received powers are considered only  in the evaluation of the association probability metric $\mathbb{P}(n=k)$.

\begin{table*}[!ht]
\centering
\caption{Comparison of Approaches for Mobility-Aware Performance Analysis in Random Single-Tier and Multi-Tier Cellular Networks}
\label{my-label}

\begin{tabular}{|p{1cm}|p{1cm}|p{2cm}|p{2cm}|p{1.5cm}|p{3cm}|p{2cm}|}
\hline
{\bf Ref} & {\bf Number of Tiers} & {\bf BS Deployment} & {\bf Mobility model}  & {\bf Approach}  & {\bf Metrics} & {\bf Association} \\ \hline
 \cite{lin2013towards} & Single-tier & Hexagonal, PPP & Modified Random Direction  & Approach-1 & Handoff rate, Sojourn time & Nearest BS
\\
\hline
\cite{bao2015stochastic}& K-tier & PPP & Arbitrary and direction $\theta \in \{0, 2\pi\}$  & Approach-1  & Handoff rate  & Biased received power
\\
\hline
\cite{adve2015handoff}& One-tier & PPP & Arbitrary and direction $\theta \in \{0, 2\pi\}$  & Approach-2  & Handoff probability, Mobility-aware coverage probability & Nearest BS
\\
\hline
\cite{shin2017equivalent} & K-tier & PPP &   Arbitrary and direction $\theta \in \{0, 2\pi\}$ & Approach-2 & Handoff probability, Mobility-aware coverage probability & Biased received power  \\ 
\hline
\cite{hesham2017velocity,heshamletter} & K-tier & PPP &   Arbitrary and direction $\theta \in \{0, 2\pi\}$ & Approach-1 & Mobility-aware throughput & Biased received power  \\ 
\hline
\end{tabular}
\end{table*}

{\cite{shin2017equivalent} provided the horizontal and vertical handoff probability expressions for  a mobile user in $K$-tier cellular network. The distance-based equivalence has been established between multiple tiers to incorporate biased received power-based association. Similar to \cite{adve2015handoff}, the coverage probability expressions consider  the  fixed handoff  cost and same mobility model.} 
The general methodology for mobility-aware coverage analysis can then be summarized as follows~\cite{shin2017equivalent}:
\begin{enumerate}
\item Derive the zero handoff probability from tier $k$ to tier $j$ using the distance-based equivalence as:
\begin{multline}
\mathbb{P}(\bar H_{k,j}|r_k, \theta)=\mathbb{P}[N(B'\backslash B'\cap A')=0|T_k \neq T_j]\\+\mathbb{P}[N(B\backslash B\cap A)=0|T_k = T_j]
\end{multline}
where $B'$ denotes the circular area centred at $l_2$ with equivalent radius of tier $j$ as $R_{j}'=w_j R_k^{-\hat{\alpha}_j}$, $A'$ denotes the circular area centred at $l_1$ with equivalent radius of tier $j$ as $r_{j}'=w_j r_k^{-\hat{\alpha}_j}$, and $r_j'$ denotes the equivalent distance of $r_j$.
\item After de-conditioning $\bar{H}_{k,j}$, the total handoff probability of a mobile user from tier $k$ can be given as:
\begin{equation}
\mathbb{P}(H_k)=1-\prod_{j=1}^{K} \mathbb{P}(\bar{H}_{k,j})
\end{equation}
Subsequently, the total handoff probability can be given as $H_0=\sum_{k=1}^K A_k \mathbb{P}(H_k)$ where $A_k$ is given by the standard association probability to tier $k$.
\item The coverage probability of a user associated to tier $k$ can then be derived using \eqref{correction}.

\end{enumerate}

{\bf Remark:} The correction in \eqref{eq:handoff_probability_modified} must also be applied to the results derived in \cite{shin2017equivalent} for multi-tier PPPs.

\subsubsection{Results}
When the network parameters are the same as in \figref{fig:Hanover_1}, the handoff rate and handoff probability for single-tier PPP derived in \cite{lin2013towards} and \cite{adve2015handoff}, and for multi-tier PPP derived in \cite{bao2015stochastic} and \cite{shin2017equivalent} are compared for low velocities in \figref{fig:Hanover_2}.
{Also, the handoff probability for single-tier PPP \cite{adve2015handoff} and for multi-tier PPP \cite{shin2017equivalent} are compared for various user velocities in \figref{fig:Hanover_3}. The analytical handoff probabilities are obtained after applying the mentioned correction to the results in \cite{adve2015handoff,shin2017equivalent}.}

\subsection{Lessons Learned}
Both the trajectory-based (Approach~1) and the association-based (Approach~2) approaches are feasible for mobility modeling in advanced cellular networks and can accommodate a variety of mobility models discussed in Section~II. The first approach can incorporate a variety of user mobility patterns with different statistics of flight length, velocity, and flight time. However, analyzing the intersections of user trajectories and the cell-boundaries may become more complicated in ultra-dense cellular networks with hot spots modeled as cluster processes. On the other hand, incorporating the impact of handoff rate on the users' coverage probability is not straight-forward. The second approach is more tractable since it characterizes handoff probability to directly evaluate coverage probability. However, the essence of user mobility models with spatial and temporal correlation may not be easy to incorporate. Sojourn time analysis was presented only in \cite{lin2013towards} using Approach~1. In the following, we provide an alternative  method for sojourn time analysis and relate it to the handoff probability analysis considering a mobile user that moves with velocity $v$ during time $T$ in a single-tier PPP cellular network. 

\subsection{Sojourn Time Analysis - Recommended Approach} We focus on sojourn time in the cell where the connection is initiated. Moreover, we only consider one movement period and no pause time. {Let us denote the duration that a mobile node stays within a particular serving cell by $S$. Then $\mathbb{P}(S>t)$ is the probability that the mobile user resides in the serving cell more than time $t$ and the sojourn time is 
\begin{equation}
\bar{S}=\mathbb{E}[S]=\int_{0}^{\infty} \mathbb{P}(S>t){\rm{d}}t.
\end{equation}
} 

{Since we are only considering one movement period, $\mathbb{P}(S>t)=0$ for $t>T$. Let us denote the location of the mobile user at time $t$ by $\text{\bf u}_t$ where $0\le t\le T$. Then $\mathbb{P}(S>t)$ can be interpreted as the probability that the mobile user is always connected to the same BS during movement from $\text{\bf u}_0$ to $\text{\bf u}_t$. Since, in single-tier PPPs with nearest BS association, the Voronoi cells are convex, when the mobile user is served by the same BS at both locations $\text{\bf u}_0$ and $\text{\bf u}_t$ we can conclude that $S>t$. Therefore, for $0\le t\le T$, $\mathbb{P}(S>t)=1-\mathbb{P}(H \mid v,t)$, where $\mathbb{P}(H \mid v,t)$ is obtained by replacing $v$ with $vt$ in \eqref{eq:handoff_probability_modified}. Finally, the sojourn time is obtained by 
	$\bar{S}=\int_{0}^{T} \mathbb{P}(S>t){\rm{d}}t=T-\int_{0}^{T} \mathbb{P}(H \mid v,t){\rm{d}}t$.}

{It is worth mentioning that we can derive sojourn time for mobility models with random velocity and random flight time by averaging $\bar{S}$ over the distribution of $v$ and $T$. Moreover, \eqref{eq:handoff_probability_modified} is derived for mobility models with uniform $\theta$. When the distribution of the movement direction is not uniform, we first need to modify \eqref{eq:handoff_probability_modified} to derive the sojourn time. It is not straightforward to apply the same approach to derive the sojourn time for multi-tier PPP cellular networks because, in multi-tier PPP cellular networks with biased nearest BS association, the Voronoi cells are not convex.}

\begin{figure}[t]
	\centering
	\includegraphics[width=\linewidth]{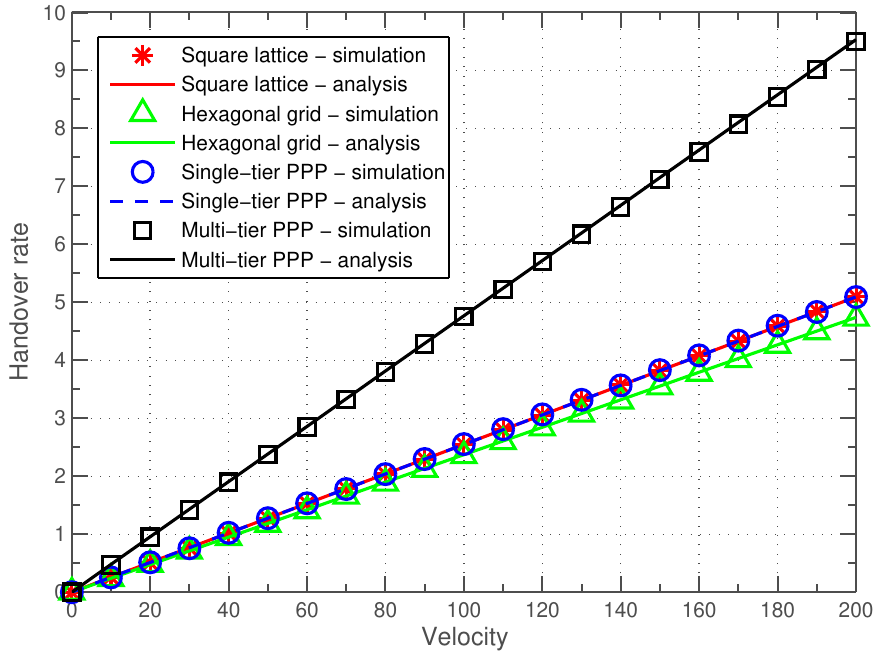}
	\caption{Handoff rate for square lattice, hexagonal grid, single-tier PPP, and multi-tier PPP with no pause time. 
		For square lattice, hexagonal grid, and single-tier PPP networks, $\lambda=0.0004$. 
		For the two-tier PPP, $\lambda_1=0.0004$, $\lambda_2=0.001$, $P_2/P_1=1/5$, $B_2/B_1=4$, and path-loss exponent $\alpha=4$.}
	\label{fig:Hanover_1}	
\end{figure}
\begin{figure}[t]
	\centering
	\includegraphics[width=\linewidth]{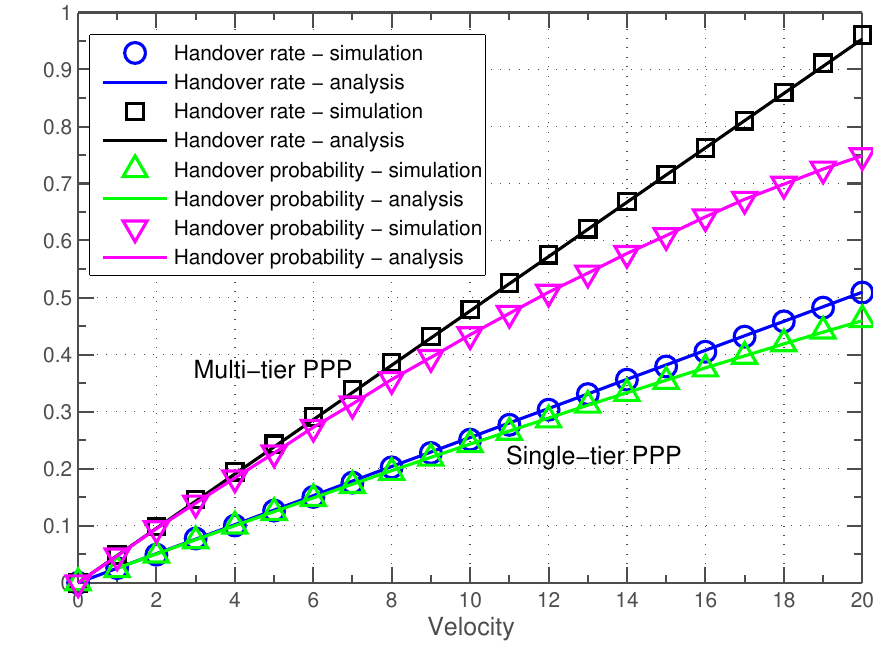}
	\caption{Handoff rate and handoff probability for low velocities. Network parameters are same as in \figref{fig:Hanover_1}.}
	\label{fig:Hanover_2}	
\end{figure}
\begin{figure}[t]
	\centering
	\includegraphics[width=\linewidth]{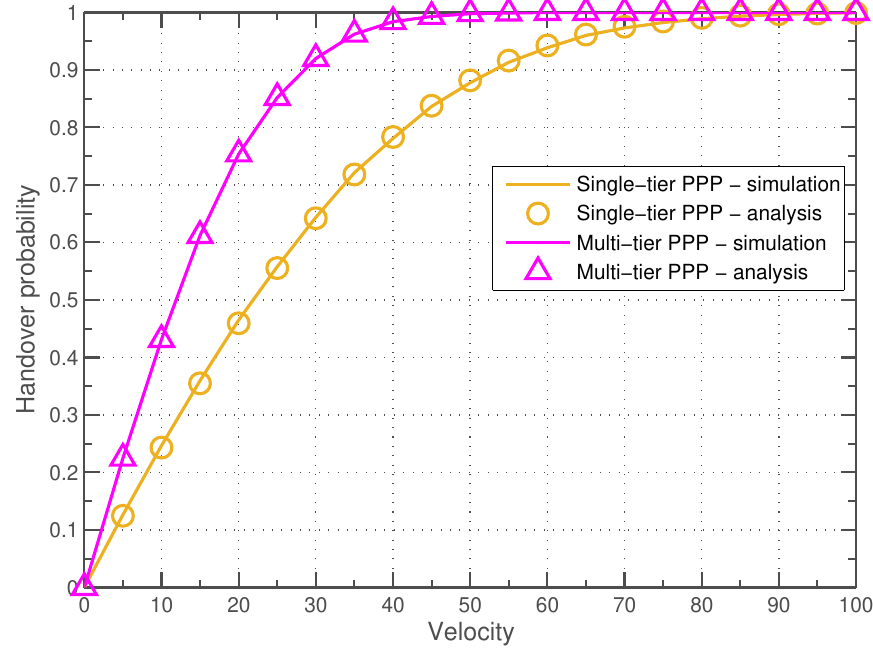}
	\caption{Handoff probability for single-tier and multi-tier PPPs. Network parameters are same as in \figref{fig:Hanover_1}.}
	\label{fig:Hanover_3}	
\end{figure}

\section{Open Challenges and Research Directions}

In this section, we will highlight the impact of mobility in emerging applications of 5G/B5G wireless cellular networks  aiming to provide  high data-rates anytime and anywhere with users/devices moving with very high-speeds (such as up to 500 km/h). Mobility management can be very challenging due to frequent  handoffs,  higher  penetration  losses, heavy  signaling  overheads,  and  sub-optimal cell selection mechanisms. Furthermore, with the emerging new wireless frequencies, software-defined resource management solutions, aerial and underwater communication, the existing  mobility  models and handoff trigger mechanisms need to be customized for specific applications.  
In the following, we discuss some of the challenges related to mobility-aware analysis and handoff management in more detail.

\subsection{Mobility-Aware Analysis of Advanced Cellular Networks}
Coverage and rate analysis for multi-tier networks assuming  homogeneous PPP deployments may not capture the correlation between  BSs and users (e.g., hot spots where more BSs or user devices tend to be installed) as well as other BSs belonging to different tiers.   It is therefore desirable to develop and analyze the user mobility  considering advanced 3GPP-inspired cellular network models~\cite{saha20173gpp} where different tiers of BSs follow Poisson cluster processes (PCP) or some tiers follow PCP and some tiers follow PPP. In addition, flexible scaling of cell sizes should be incorporated to consider advanced user association mechanisms such as biased user association, channel access aware user association, backhaul aware association, decoupled uplink-downlink association, etc. Clearly, for such scenarios, the expressions for handoff rates experienced by an active mobile user with arbitrary movement trajectory will be novel and may even be complicated. Mobility-aware handoff, coverage and rate analysis for aforementioned models may also choose to incorporate higher frequencies such as optical frequencies.  In addition, three dimensional mobility models based on stochastic geometry taking into account the antenna height, azimuth, and beam-forming for drone applications are of immediate relevance.

\subsection{Mobility in Millimeter-Wave and Nanometer-Wave Networks}
A fundamental challenge in high frequency (e.g., millimeter-wave [mm-wave] frequency, nanometer-wave [nm-wave] or optical frequency) transmission is intermittent connectivity due to blocking, channel attenuation,  and availability of only line-of-sight (LoS) propagation. The signal strength degradation as a function of distance results in a small coverage area, thereby causing frequent handoffs. Contrary to traditional radio frequencies, small movements of obstacles, reflectors, and changes in the orientation of a handset relative to the body or a hand, can vary the channel rapidly. Thus a channel handoff may even be required while associated to the same BS and with very low mobility. It is thus crucial to exploit macro-diversity, i.e., alternate channels should be available for rapid reconnection. For example, due to higher isotropic path-loss, mm-wave transmissions are always supported with phased arrays or massive antennas; thus signals are transmitted in narrow beams.  As such, frequent handoffs (even for users with low mobility levels) are a big concern.  Mobility-aware analysis of such networks as well as efficient mobility management in these networks are open research problems.  


To overcome the higher handoff rates and handoff cost in these networks (due to small cell coverage areas), intelligent handoff skipping techniques would be essential to maintain a trade-off between handoff rate and throughput. For example, the decision of a user to skip the BS may be based on the coverage area of the cell, the velocity of the user, trajectory distance or dwell time within the cell, and the interference and transmission frequency offered by the BS. Also,  the handoff skipping decisions can be based on the traffic load conditions of various consecutive BSs on the user's trajectory as well as the velocity patterns of nearby users.  It was shown that handoff skipping strategies are an effective way to minimize handoff rates while improving the QoS requirements of the users and this way achieving a balance between the two~\cite{skipping1, skipping2,hesham2017velocity}.

\subsection{Mobility-Aware Analysis of Wireless Energy Transfer/Harvesting Networks}
Compared to wireless information transfer, wireless energy transfer is relatively more vulnerable to the mobility of a user~\cite{galinina2016feasibility}. The reason is that the desired receiver sensitivity for wireless energy harvesting (reported to be -10 dBm) is higher than the desired receiver sensitivity for data receptions (which is around -60 dBm). Due to different sensitivity levels, the successful energy harvesting distance can be noticeably low. Consequently, depending on the level of mobility of users, the number of handoffs for wireless charging can be
significantly higher. As such, sophisticated analytical models would be required that can model the velocity of mobile energy harvesting devices and characterize a precise mobility-aware handoff criteria for efficient energy transfer.
In a multi-operator environment,  different
operators will likely  have different energy levels, spectrum, traffic load, and operational cost. 
For instance, given the strong energy transfer capabilities
of a specific operator, other operators may achieve higher
incentives by off-loading their associated energy harvesting
devices to the strong operator. In return, the strong operators
(i.e., in terms of energy transfer capability) can either demand
incentives or offload their users to other nearby operators
which are stronger from the perspective of wireless resources
and information transfer. New analytical/optimization frameworks should thus be developed to model and evaluate
the energy trading process in wireless energy harvesting networks.

\subsection{Mobility in UAV networks}
Compared to terrestrial networks, UAV/drone networks have their unique attributes such as the mobility of both transmitter and receivers, altitude of the drones, frequent network topology changes, mechanical and aerodynamic constraints, strict safety requirements, and harsh communication in the disconnected, intermittent, limited bandwidth (DIL) environments. Moreover, field tests in air and emulations might be very costly and scenario-specific  to assess the handoff performance. To cope with this, simulations using random mobility models are generally a low-cost, systematic and robust alternative.   However, the traditional random mobility models  may not correctly emulate aerial networks.  A redesign is therefore necessary in order to develop reliable simulation environment and subsequent design/evaluation of the drone networks. 
Compared to air-to-air (AtA), an air-to-ground (AtG) handoff procedure may also be vital  in ensuring efficient network resource consumption while attaining QoS requirements of the users. Since drones in near future may likely deploy WiFi due to lighter payload compared to LTE~\cite{uav7}, it may be difficult to perform seamless handoff because a traditional WiFi network not only has a narrow communication coverage  but also gives a relatively long handoff time \cite{uav8}. Traditional handoff decision algorithms typically assume that the coverage of each BS is the same, which does not apply to drone networks with different size, weight, and power (SWAP) constraints and altitudes in the three-dimensional (3-D) space \cite{uav9}. Based on the information of the received signal strength (RSS), drones can potentially adjust their speed and height as well as distance from other drones. 

\subsection{Mobility in Software-Defined Cellular Networks}
Mobility management has evolved significantly from managing  single-RAT to multi-RAT networks. Nevertheless, network slicing in 5G  will bring  mobility related challenges due to heterogeneity in the characteristics, latency, and reliability requirements of various slices. Ultra-reliable and low latency communication (uRLLC) slice, Internet-of-things (IoT) slice and enhanced mobile broadband (eMBB) slices are three main categories of network slicing in 5G. For example, in uRLLC slicing, communication devices are more sensitive to time delay and require lower transmission rate than those in other slices. Communication requirements in air planes or high-speed trains are expected to trigger significant handoffs compared to IoT slices with static or nearly static devices. Therefore, the handoff procedure may change significantly among various network slices. Note that, traditionally handoffs are event-triggered. However, in network slicing-based 5G networks, flexible handoff criteria and adaptive handoff thresholds are needed to support mobility management in SDN. Multi-RAT coordination may also be exploited among different RATs to share location information of their respective mobile devices. \cite{hesham2016split} analyzed the negative impact of user mobility  in dense  multi-tier  cellular  networks and  the  concept of  Phantom  small  BSs (split control and data plane architecture)  has been advocated  to  mitigate  such  negative impact. Furthermore, the handoff delay due to handoff related signaling can be tackled using SDN paradigm. For example, a mobile node that  exists in one of the BSs may have a number of  neighboring BSs and thus have finite transition probability associated with each neighboring BS. The controller can  utilize the automatic neighbor relation  (ANR) function of the BSs. With the  neighbor removal and detection functions of the ANR, the neighbor relation tables and transition probabilities can be constructed.



\subsection{Mobility-Aware C/U Plane Splitting}
With decoupled control and data plane,  a BS with wider coverage handles control plane signaling while BSs with smaller coverage perform data plane transmissions. With this decoupled control and data plane architecture, the mobile users' experience becomes robust since the number of required handoffs will reduce significantly. This will end up vacating more spectrum resources at the BSs with smaller coverage zones due to reduced control information and reference signals transmission from small cells. Clearly, the control signaling with macrocell BSs would be much advantageous for fast moving users as it can efficiently eliminate the possibility of frequent handoffs, thus significantly reducing the related signaling overhead.  Note that if the control connection is established with  small cells, the handoff procedure may not be completed before the user moves out of the coverage of the target cell. However, for low-mobility scenarios,  the control plane handoffs can be established with small cells. There is almost no need for frequent handoff and the shorter access distance to the nearest small cell will naturally lead to enhanced energy efficiency. 


\subsection{Mobility-Aware Caching}
With the increase in mobile traffic and repeated content requests, caching at the BSs becomes important to
reduce the excessive backhaul load and delay,  particularly  for  on-demand  video-streaming  applications. Generally, it is assumed that the user can download the entire requested content through the connected BSs. However, it is noteworthy that the user mobility in such a scenario makes the contact duration between user and BSs random which  directly impacts the cache hit and miss probabilities. Subsequently, the content placement and the content delivery strategy should take mobility  into account. A few recent works have has addressed this issue~\cite{cache1}. 


\subsection{Data-Driven Learning Models for Mobility-Aware Analysis}
For a dense, multi-tier, and heterogeneous 5G/B5G  network, in some scenarios, it may not be possible to accurately model the mobility of users/devices to analyze system metrics such as the handoff rate, cell sojourn time, mobility-aware coverage and rate.  In these scenarios, data-driven machine learning models~\cite{learningsurvey1,learningsurvey2} such as the deep neural networks (DNNs) may serve as an important tool to predict the system parameters and network performance.  The essence of data-driven machine learning is, a pattern exists in the system behavior which cannot be pinned down mathematically. However, data is available and the system behavior can be learnt (or predicted) from the input data set by analyzing the data. Deep learning (DL) is a branch of machine learning and a DNN is a neural network (NN)-based DL architecture consisting of the input and output layers and many hidden layers of {\em  neurons}, which can perform non-linear functions. The neurons in a hidden layer  are fully-connected and weights are assigned to various neurons. 
The auto-encoder NN, convolutional NN, and recurrent NN are most common types of DNNs. A DNN is trained by a subset of available data called the training set and then tested by using another set of data called the test data. A DNN can be used to learn the mobility patterns of network nodes, and based on this, network performance metrics can be predicted. 
Nevertheless, in order to achieve high accuracy, a significant amount of input data may be required and also the DNN will need to be optimized (e.g., in terms of the number of hidden layers and number of neurons per layer). 

\bibliography{Reference}
\bibliographystyle{IEEEtran}


\end{document}